\documentclass[11pt,a4paper]{article}
\pdfoutput=1

\usepackage{jheppub}
\usepackage[dvipsnames]{xcolor}
\usepackage{mathrsfs}
\usepackage{amsmath,amsthm,amssymb,slashed}
\usepackage{color}
\usepackage{multirow}
\usepackage[normalem]{ulem}
\usepackage{afterpage}
\usepackage{url}
\usepackage{hyperref}

\newcommand{\be}{\begin{equation}}
\newcommand{\ee}{\end{equation}}
\newcommand{\bea}{\begin{eqnarray}}
\newcommand{\eea}{\end{eqnarray}}
\newcommand{\ba}{\begin{array}}
\newcommand{\ea}{\end{array}}

\newcommand{\Qwq}{\mathcal{Q}^2}

\newcommand{\dresden}{Dresden-II }
\newcommand{\dd}{\mathrm{d}}
\newcommand{\Eps}{\varepsilon}

\allowdisplaybreaks

\title{Bounds on new physics with data of the Dresden-II reactor experiment and COHERENT}

\author[a]{Pilar Coloma,}
\affiliation[a]{Instituto de F\'isica Te\'orica UAM-CSIC, Universidad Aut\'onoma de Madrid, \\ Calle de Nicol\'as Cabrera 13--15, Cantoblanco, E-28049 Madrid, Spain}
\emailAdd{pilar.coloma@ift.csic.es}

\author[b,c]{Ivan Esteban,}
\affiliation[b]{Center for Cosmology and AstroParticle Physics (CCAPP), Ohio State University, \\
	191 W. Woodruff Ave., Columbus, Ohio 43210, U.S.A.}
\affiliation[c]{Department of Physics, Ohio State University, \\ 
	191 W. Woodruff Ave., Columbus, Ohio 43210, U.S.A.}
\emailAdd{esteban.6@osu.edu}

\author[d,e,f]{M.C. Gonzalez-Garcia,}
\affiliation[d]{C. N. Yang Institute for Theoretical Physics, Stony Brook University, \\ 
	Stony Brook NY 11794-3840, U.S.A.}
\affiliation[e]{Institució Catalana de Recerca i Estudis Avançats (ICREA), \\
	Pg.\ Lluis Companys 23, E-08010 Barcelona, Spain}
\affiliation[f]{Departament de Fisica Quantica i Astrofisica and Institut de Ciencies del Cosmos, \\ Universitat de Barcelona, Diagonal 647, E-08028 Barcelona, Spain}
\emailAdd{maria.gonzalez-garcia@stonybrook.edu}

\author[g]{\\ Leire Larizgoitia,}
\affiliation[g]{Donostia International Physics Center (DIPC), \\ 
	Paseo Manuel Lardizabal, 4, Donostia-San Sebasti\'an, E-20018, Spain}
\emailAdd{leire.larizgoitia@dipc.org}

\author[g,h]{Francesc Monrabal,}
\affiliation[h]{Ikerbasque, Basque Foundation for Science, \\ 
	Plaza Euskadi 5, E-48013 Bilbao, Spain}
\emailAdd{francesc.monrabal@dipc.org}

\author[i]{Sergio Palomares-Ruiz}
\affiliation[i]{Instituto de F\'isica Corpuscular (IFIC), CSIC \& Universitat de València, \\
  Parc Científic, C/ Catedr\'atico Jos\'e Beltr\'an 2, E-46980 Paterna, Spain}
\emailAdd{sergiopr@ific.uv.es}

\abstract{
Coherent elastic neutrino-nucleus scattering was first experimentally established five years ago by the COHERENT experiment using neutrinos from the spallation neutron source at Oak Ridge National Laboratory. The first evidence of observation of coherent elastic neutrino-nucleus scattering with reactor antineutrinos has now been reported by the Dresden-II reactor experiment, using a germanium detector. In this paper, we present constraints on a variety of beyond the Standard Model scenarios using the new Dresden-II data. In particular, we explore the constraints imposed on neutrino non-standard interactions, neutrino magnetic moments, and several models with light scalar or light vector mediators. We also quantify the impact of their combination with COHERENT (CsI and Ar) data. In doing so, we highlight the synergies between spallation neutron source and nuclear reactor experiments regarding beyond the Standard Model searches, as well as the advantages of combining data obtained with different nuclear targets. We also study the possible signal from beyond the Standard Model scenarios due to elastic scattering off electrons (which would pass selection cuts of the COHERENT CsI and the Dresden-II experiments) and find more stringent constraints in certain parts of the parameter space than those obtained considering coherent elastic neutrino-nucleus scattering.}

\preprint{\\[-38pt]\begin{flushright}IFT-UAM/CSIC-22-10 \\ YITP-SB-2022-03 \\ IFIC/22-06 \end{flushright}}


\begin{document}

\maketitle

\section{Introduction}
\label{sec:intro}

Low-energy neutrinos can elastically scatter off atomic nuclei via weak neutral currents, with the initial and final states of the nucleus being indistinguishable. For low enough momentum transfers, the interaction takes place coherently with the whole nucleus, leading to an enhancement of the cross section approximately proportional to the square of its number of neutrons. Although the so-called coherent elastic neutrino-nucleus scattering (CE$\nu$NS) was first theoretically described in Freedman's seminal paper almost 50 years ago~\cite{Freedman:1973yd}, its detection has not been possible until very recently. This is so because the single observable for this process is a recoiling nucleus which generates a signal in the sub-keV to few keV energy range, difficult to detect. An additional hindrance to CE$\nu$NS detection is the limited number of suitable neutrino sources, which must be sufficiently intense in yield and, at the same time, low enough in neutrino energy for the coherence condition to be satisfied.

The detection of CE$\nu$NS was experimentally demonstrated by the COHERENT experiment~\cite{COHERENT:2017ipa} using the currently most intense spallation neutron source (SNS), sited at the Oak Ridge National Laboratory, U.S.A. The original observation was obtained with a CsI[Na] scintillation detector~\cite{COHERENT:2017ipa, COHERENT:2018imc}, which was later followed by the observation of CE$\nu$NS at a liquid Argon detector~\cite{COHERENT:2020iec, COHERENT:2020ybo}.

In addition to spallation sources, CE$\nu$NS is also searched for in a variety of experiments using electron antineutrinos emitted by nuclear reactors, such as TEXONO~\cite{Wong:2004ru}, $\nu$GeN~\cite{Belov:2015ufh}, CONNIE~\cite{Aguilar-Arevalo:2016khx}, MINER~\cite{Agnolet:2016zir}, Ricochet~\cite{Billard:2016giu}, $\nu$-cleus~\cite{Strauss:2017cuu}, RED-100~\cite{Akimov:2019ogx}, NEON~\cite{Choi:2020gkm}, CONUS~\cite{CONUS:2020skt} and NCC-1701 at \dresden~\cite{Colaresi:2021kus}. At present, most of these reactor experiments have not yet been successful in the CE$\nu$NS detection. The exception to this is the NCC-1701 experiment at the \dresden nuclear reactor. In their first published data~\cite{Colaresi:2021kus}, the experimental collaboration presented an event spectrum with an excess of events at low energies, compatible with expectations from a CE$\nu$NS signal in the Standard Model (SM). Recently, the collaboration has released new results with an increased exposure. The observation of CE$\nu$NS is reported with strong to very strong preference (with respect to the only-background hypothesis) in the Bayesian statistics sense, depending on the quenching factor considered~\cite{Colaresi2022suggestive}.

A precision measurement of CE$\nu$NS provides a direct probe to both SM and beyond the standard model (BSM) physics. Paradigmatic examples of the former are the determination of the weak mixing angle at very low momentum transfer~\cite{Canas:2018rng, Cadeddu:2018izq, Huang:2019ene, Cadeddu:2019eta, Cadeddu:2020lky, Cadeddu:2021ijh} and the study of nuclear structure~\cite{Cadeddu:2017etk, Ciuffoli:2018qem, Papoulias:2019lfi, Cadeddu:2021ijh, Coloma:2020nhf}. The program of BSM exploration with CE$\nu$NS is broad (see, e.g., refs.~\cite{Barranco:2005yy, Formaggio:2011jt, Anderson:2012pn, Dutta:2015nlo, Cerdeno:2016sfi, Dent:2016wcr, Coloma:2017egw, Kosmas:2017zbh, Ge:2017mcq, Shoemaker:2017lzs, Coloma:2017ncl, Liao:2017uzy, Canas:2017umu, Dent:2017mpr, Papoulias:2017qdn, Farzan:2018gtr, Billard:2018jnl, Coloma:2019mbs, Chaves:2021pey, AristizabalSierra:2018eqm, Brdar:2018qqj, Cadeddu:2018dux, Blanco:2019vyp, Dutta:2019eml, Miranda:2019wdy, CONNIE:2019swq, Dutta:2019nbn, Papoulias:2019txv, Khan:2019cvi, Cadeddu:2019eta, Giunti:2019xpr, Baxter:2019mcx, Canas:2019fjw, Miranda:2020zji, Flores:2020lji, Miranda:2020tif, Hurtado:2020vlj, Miranda:2020syh, Cadeddu:2020nbr, Shoemaker:2021hvm, delaVega:2021wpx, Liao:2021yog, CONUS:2021dwh, Flores:2021kzl, Li:2022jfl, AristizabalSierra:2019zmy, Abdullah:2020iiv, Fernandez-Moroni:2021nap, Bertuzzo:2021opb, Bonet:2022imz} for an incomplete list) being most sensitive to a variety of scenarios leading to modified neutrino interactions with nuclei -- in particular at low momentum transfer -- but extending also to the production of new light neutral states, and sterile neutrino searches, among others.

In this paper, we present the first analysis using the new data of the \dresden reactor experiment in the context of BSM searches. In doing so, we consider several new physics scenarios commonly studied in the literature: effective four-fermion interactions, neutrino magnetic moments, and light mediators. In order to place our results in a broader context, we also combine the \dresden data with a germanium detector with the COHERENT data obtained for neutrino scattering off CsI and Ar. We highlight and quantify the synergies between SNS and reactor experiments using CE$\nu$NS, as well as the advantages of the combination of data sets obtained with multiple nuclear targets. Furthermore, we also study a potential signal from BSM scenarios due to elastic electron scattering (ES). Although in the SM this contribution to the event rates is negligible, it could be significantly enhanced in the presence of new physics effects. The contribution from scattering off electrons could pass selection cuts of both COHERENT CsI and \dresden experiments and, as we discuss, its inclusion leads to stronger constraints in certain parts of the parameter space than those obtained using CE$\nu$NS.

The paper is organized as follows. In section~\ref{sec:frameworks} we introduce the theoretical frameworks we consider. Section~\ref{sec:fit} discusses the computation of the expected event rates and the details regarding our fit to the experimental data, while the results of our analysis are presented in section~\ref{sec:results}. Finally, we summarize and conclude in section~\ref{sec:conclusions}, and provide the binding energies for neutrino scattering off electrons in appendix~\ref{app:levels}.

\section{Phenomenological frameworks}
\label{sec:frameworks}

\begin{table}
	\begin{center}
		\begin{tabular}{|c|c|c|}
			\hline
			& & \\[-13pt]
			& $c_V$ & \; \quad \; $c_A$ \; \quad \; \\[1pt] \hline
			& & \\[-9pt]
			\; $\nu_e$ \; \quad & \; $\frac{1}{2} + 2 \, \sin^2\theta_{W}$ \; & \; $+\frac{1}{2}$ \; \\[8pt]
			\; $\nu_\mu, \nu_\tau$ \; \quad & \; $-\frac{1}{2} + 2 \, \sin^2\theta_{W}$ \; & \; $-\frac{1}{2}$ \; \\[3pt] \hline
		\end{tabular}
	\end{center}\vspace{-0.2cm}
	\caption{\label{tab:cs} Values of the SM effective couplings in neutrino-electron scattering. For antineutrinos the vector couplings are the same, while the axial ones change sign, $c_A \to - c_A$.}
\end{table}

Let us start by introducing the differential cross sections for both CE$\nu$NS and ES in the SM. The SM differential coherent elastic scattering cross section, for a neutrino with energy $E_\nu$ scattering off a nucleus of mass $m_A$, is given\footnote{This strictly applies to $J=1/2$ nuclei~\cite{Lindner:2016wff}. Note that for $J=0$ nuclei the last term is not present~\cite{Freedman:1973yd}, although its contribution is negligible for the energies of interest.} by~\cite{Freedman:1973yd, Lindner:2016wff}
\begin{equation}
\label{eq:sm-nucl}	
\frac{\dd \sigma^{\rm CE\nu NS}_{\rm SM}}{\dd E_R} = {\cal Q}_W^2 \,
\frac{G_F^2 \, m_A}{2 \, \pi} \, \left[ 2 - \frac{m_A \, E_R}{ E_{\nu}^2} - \frac{2 \, E_R}{E_{\nu}} + \left( \frac{E_R}{E_{\nu}} \right)^2 \right] \, \left|F(q)\right|^2 ~,
\end{equation}
where $G_F$ is the Fermi constant, $E_R$ is the nuclear (kinetic) recoil energy, and the weak hypercharge of the nucleus in the SM is ${\cal Q}_{W} = \left( N \, g_{V, n} + Z \, g_{V,p} \right)$, where $N$ and $Z$ are its number of neutrons and protons, respectively. The weak charges of the neutron and the proton are $g_{V, n} = -1/2$ and $g_{V,p} = 1/2 - 2 \sin^2 \theta_W$, with $\sin^2 \theta_W = 0.23868$ at zero momentum transfer~\cite{Erler:2004in, Erler:2017knj}. For Ge we use $\{N, Z\} = \{40.71, \, 32\}$ (weighted average of natural isotopic abundances), while for CsI we use $\{N, Z\} = \{76, \, 54\}$ and for Ar, $\{N, Z\} = \{22, \, 18\}$.

In eq.~\eqref{eq:sm-nucl} above, the nuclear form factor $F(q)$ is assumed to be equal for protons and neutrons. Some commonly used phenomenological parameterizations are the Klein-Nystrand~\cite{Klein:1999qj} or the Helm~\cite{Helm:1956zz} form factors, among others. Note that, at the energies of interest for the \dresden experiment, $F(q) \simeq 1$, so the choice of form factor is completely irrelevant. For COHERENT, we use two different form factors, depending on the target nucleus. For CsI, we use the theoretical calculation from refs.~\cite{Klos:2013rwa} (based on refs.~\cite{Hoferichter:2018acd, Hoferichter:2016nvd}), while for Ar we assume a Helm form factor with $s = 0.9 ~\mathrm{fm}$, $\langle r_n^2\rangle = -0.1161 ~\mathrm{fm}^2$ and $\langle r_p^2\rangle = 0.70706~\mathrm{fm}^2$~\cite{ParticleDataGroup:2020ssz}, $R_\mathrm{ch} = 3.4274~\mathrm{fm}$~\cite{Angeli:2013epw} and $R_{n,\mathrm{pt}} = 3.405~\mathrm{fm}$~\cite{Payne:2019wvy} (following the same notation as in ref.~\cite{Coloma:2020nhf}).

The differential cross section for ES per atom within the SM can be expressed as (see, e.g., refs.~\cite{Vogel:1989iv, Tomalak:2019ibg})
\begin{equation}
\label{eq:sm-elec}
\frac{\dd\sigma^{\rm ES}_{\rm SM}}{\dd T_e} = Z_{\rm eff}^{\rm X}(T_e) \, \frac{G_F^2 \, m_e}{2 \, \pi} \left[ \left(c_V + c_A\right)^2 + \left(c_V - c_A\right)^2\left(1-\frac{T_e}{E_\nu} \right) + \left(c_A^2 - c_V^2\right)\frac{m_e \, T_e}{E_\nu^2} \right] ~ ,
\end{equation}
where $T_e$ stands for electron (kinetic) recoil energy. The effective electron couplings $c_V$ and $c_A$ contain the contributions from the SM neutral current (NC), which are equal for all flavors, plus the charged-current (CC) contribution which only affects $\bar\nu_e e^-$ and $\nu_e e^-$ scattering. For convenience, their values are given in table~\ref{tab:cs}.

We assume that scattering off electrons is incoherent, so the total ES cross section per atom is obtained by multiplying the ES cross section per electron by an effective charge, $Z_{\rm eff}^{\rm X}(T_e)$, where $\textrm{X} \equiv ^A_Z\!\text{X}$ indicates the target nucleus. This takes into account that, in this process, atomic binding effects have to be considered for recoil energies comparable to atomic binding energies. In doing so, we follow the procedure proposed in refs.~\cite{Kopeikin:1997, Fayans:2000ns, Mikaelyan:2002nv} and assume $Z_{\rm eff}^{\rm X}(T_e)$ to be a step function of the binding energies of the different atomic levels, as provided in appendix~\ref{app:levels}.

Finally, both types of interactions (off nuclei and off electrons) are two-body elastic scattering processes, so the kinematics is the same. Hence, the minimum neutrino energy to produce a recoil with energy $T$ is $E_{\nu, {\rm min}} = \left( T + \sqrt{T^2 + 2 \, m \, T}\right)/2$ and the maximum recoil energy for a given neutrino energy is $T_{\rm max} = 2 \, E_\nu^2/(2 \, E_\nu + m)$, where $T \equiv E_R$ and $m \equiv m_A$ for nuclei, while $T \equiv T_e$ and $m \equiv m_e$ for electrons.

Throughout this work we will consider several phenomenological scenarios leading to modifications to the interaction cross sections in eqs.~\eqref{eq:sm-nucl} and~\eqref{eq:sm-elec}, as explained in more detail below.

\subsection{Non-standard neutrino interactions}
\label{sec:fraNSI}

The so-called non-standard neutrino interaction (NSI) framework consists on the addition of four-fermion effective operators to the SM Lagrangian at low energies. For example, the effective Lagrangian
\begin{equation}
\label{eq:nsi-nc}
\mathcal{L}_\text{NSI, NC} = - \sum_{f,\alpha,\beta} 2\sqrt{2} \, G_F \, \Eps_{\alpha \beta}^{f,P} \, \left(\bar\nu_\alpha \gamma_\mu P_L \nu_\beta\right) \, \left(\bar f \gamma^\mu P f\right) ~, 
\end{equation}
would lead to new NC interactions with the rest of the SM fermions. Here, $\alpha,\beta\equiv e, \mu, \tau$ while $f$ refers to SM fermions, and $P$ can be either a left-handed or a right-handed projection operator ($P_L$ or $P_R$, respectively).  Such new interactions may induce lepton flavor-changing processes (if $\alpha \neq \beta$), or may lead to a modified interaction rate with respect to the SM result (if $\alpha = \beta$).

In presence of NC NSI, the effective charge of the nucleus in eq.~\eqref{eq:sm-nucl} gets modified, $\mathcal{Q}_W^2 \to \Qwq_\alpha(\boldsymbol{\Eps})$. For real off-diagonal NSI parameters, it reads~\cite{Barranco:2005yy}
\begin{eqnarray}
\label{eq:Qalpha-nsi}
\Qwq_\alpha(\boldsymbol{\Eps}) & = & \left[ Z \big(g_p^V + 2 \, \Eps_{\alpha\alpha}^u + \Eps_{\alpha \alpha}^d \big) + N \big( g_n^V + \Eps_{\alpha\alpha}^u + 2 \, \Eps_{\alpha \alpha}^d \big)  \right]^2 \nonumber \\ 
& & + \sum_{\beta \neq \alpha} \left[ Z \big( 2 \, \Eps_{\alpha \beta}^u + \Eps_{\alpha \beta}^d) + N \big( \Eps_{\alpha\beta}^u + 2 \, \Eps_{\alpha\beta}^d \big) \right]^2 ~.
\end{eqnarray}
where, in order to simplify notation, we have renamed $\Eps_{\alpha\beta} \equiv \Eps_{\alpha\beta}^{q,V} = \Eps_{\alpha\beta}^{q,L} + \Eps_{\alpha\beta}^{q,R}$. The weak charge may be rewritten in a more compact form as
\begin{eqnarray}
\label{eq:Qalpha-nsi-X}
\Qwq_{\alpha}(\boldsymbol{\Eps}) & = & \left[ \mathcal{Q}_{W} + \Eps_{\alpha\alpha}^\text{X} \right]^2 + \sum_{\beta\neq \alpha} \big( \Eps_{\alpha\beta}^\text{X} \big)^2 ~,
\end{eqnarray}
where we have defined
\begin{equation}
\label{eq:eps-nucleon}
\Eps_{\alpha\beta}^\text{X} \equiv N \, \Eps_{\alpha\beta}^n + Z \, \Eps_{\alpha\beta}^p \,, \qquad \Eps_{\alpha\beta}^n \equiv \Eps_{\alpha\beta}^u + 2 \, \Eps_{\alpha\beta}^d ~, \qquad \Eps_{\alpha\beta}^p \equiv 2 \, \Eps_{\alpha\beta}^u + \Eps_{\alpha\beta}^d ~.
\end{equation}

The first consequence we observe of the inclusion of NSI effects is that the weak charge may now depend on the incident neutrino flavor $\alpha$. As will be discussed below, it is relevant to note that the COHERENT experiment observes interactions of both electron and muon neutrinos. However, at first approximation, it measures the linear combination $\Qwq_e + 2 \, \Qwq_\mu$ and hence, both charges are degenerate: a reduction in the value of $\Qwq_e$ can be compensated by an increase in $\Qwq_\mu$ and vice versa. At COHERENT, though, the addition of timing information partially breaks this degeneracy for CsI, as discussed in detail in ref.~\cite{Coloma:2019mbs}. Including reactor data, which is only sensitive to interactions with electron antineutrinos, brings in additional complementary information in this respect~\cite{Dent:2017mpr}. It provides an additional constraint on $\Qwq_e$, which is independent of $\Qwq_\mu$.

The second relevant feature we observe from eqs.~\eqref{eq:Qalpha-nsi}-\eqref{eq:eps-nucleon} is that the impact of NSI on the weak charge depends on the values of $N$ and $Z$ in a non-trivial manner. Because of this, the combination of data obtained for different nuclei offers an additional handle to reduce the size of the allowed confidence regions of this scenario (for earlier discussions see, e.g., refs.~\cite{Scholberg:2005qs, Barranco:2005yy, Coloma:2017egw, Baxter:2019mcx}).

\subsection{Neutrino magnetic moment}
\label{sec:framag}

In the presence of a neutrino magnetic moment, $\mu_\nu$, the cross sections for neutrino scattering off nuclei and electrons get additional contributions which do not interfere with the SM ones. The scattering off protons can be considered coherent and therefore its cross section is given, up to order ${\cal O}((E_R/E_\nu)^2, E_R/m_A)$, by~\cite{Vogel:1989iv}
\begin{equation}
\label{eq:mag-nuc}	
\frac{\dd\sigma^{\rm CE\nu NS}_{\rm \mu_\nu}}{\dd E_R} = Z^2 \,
\left(\frac{\mu_\nu}{\mu_B}\right)^2 \, \frac{\alpha^2\,\pi}{m_e^2}
\left[\frac{1}{E_R} - \frac{1}{E_\nu}\right]\; |F(q)|^2 ~,
\end{equation}
where the form factor $F(q)$ is assumed to be the same as in the SM,  which is a reasonable approximation at the transfer momenta of interest~\cite{Hoferichter:2020osn}. 

Conversely, the scattering off electrons is incoherent and therefore, the cross section in this case, up to order ${\cal O}((T_e/E_\nu)^2)$, is~\cite{Vogel:1989iv}
\begin{equation}
\label{eq:mag-elec}	
\frac{\dd\sigma^{\rm ES}_{\rm \mu_\nu}}{\dd T_e} = Z_{\rm eff}^{\rm X}(T_e) \, \left(\frac{\mu_\nu}{\mu_B}\right)^2 \, \frac{\alpha^2\,\pi}{m_e^2} \left[\frac{1}{T_e}-\frac{1}{E_\nu}\right] ~.
\end{equation}
Notice that, in writing eqs.~\eqref{eq:mag-nuc} and~\eqref{eq:mag-elec}, we have denoted the neutrino magnetic moment as $\mu_\nu$, without specifying the flavor of the neutrino. However, neutrino magnetic moments arise in a variety of models of new physics and, in particular, they do not need to be flavor-universal. Therefore, in what follows, we will allow different magnetic moments for the different neutrino flavors, reporting their bounds separately.

\subsection{Light scalar mediators}
\label{sec:frasca}

The Lagrangian of the simplified model of interaction of a scalar $\phi$ with the relevant fermions we consider is
\begin{equation}  
\label{eq:lagscal}    	
{\cal L}_{\phi} = g_\phi \, \phi \, \left( \sum_q q_\phi^q \, \bar{q}q + q_\phi^e\, \bar{e} e + q_\phi^\nu\, \bar{\nu}_{R} \nu_{L} + \text{h.c.} \right) - \frac{1}{2} \, M^2_\phi \, \phi^2 ~,
\end{equation}
where $q_\phi^i$ are the charges of each fermion ($i = \{\nu,\, e,\, u,\, d,\, s,\, c,\, b,\, t\}$) under the new interaction. This new interaction does not interfere with the SM one~\cite{Rodejohann:2017vup, Farzan:2018gtr}, and the corresponding neutrino-nucleus elastic scattering cross section, up to order ${\cal O}((E_R/E_\nu)^2)$, reads~\cite{Cerdeno:2016sfi, Farzan:2018gtr}
\begin{equation}
\frac{\dd\sigma^{\rm CE\nu NS}_\phi}{\dd E_R} = \frac{g_\phi^4\, (q_\phi^\nu)^2 \, \mathcal{Q}_\phi^2}{4 \, \pi} \, \frac{m_{A}^2 \, E_R }{E_{\nu}^2 \, (2 \, m_{A} \, E_R + M_\phi^2)^2} \, |F(q)|^2 ~,
\label{eq:csscaln}
\end{equation}
where $F(q)$ is the form factor (which, with enough precision at the transfer momenta of interest, can be assumed to be the same as in the SM~\cite{Hoferichter:2020osn}), and $\mathcal{Q}_\phi$ is the nuclear charge for the scalar interaction~\cite{Shifman:1978zn, DelNobile:2013sia, Ellis:2018dmb},
\begin{equation}
\label{eq:qscal}
\mathcal{Q}_{\phi} = \sum_{N, q} q^q_\phi \, \frac{m_N}{m_q} \, f^{(N)}_{T,q} + \frac{2}{27} \, \left( 1 - \sum_{N, q} f^{(N)}_{T,q} \right) \sum_{N, \tilde{q}} q_\phi^{\tilde{q}} \, \frac{m_N}{m_q} ~.
\end{equation}
Here $N=n,p$ stands for the nucleons, the superindex $q=u,d,s$ runs over the light valence and sea quarks, $\tilde{q}=c,b,t$ runs over the heavy quarks, and the coefficients $f_{T,q}^{(N)}$ incorporate the effective low-energy couplings of the scalar to the nucleons~\cite{Shifman:1978zn}. For universal coupling of the scalar to quarks, $q_\phi^u = q_\phi^d = q_\phi^s = q_\phi^c = q_\phi^b = q_\phi^t \equiv q_\phi^q$, the nuclear scalar charge takes the value $\mathcal{Q}_{\phi} = q_\phi^q \, (14 \, N + 15.1 \, Z)$~\cite{DelNobile:2013sia}.\footnote{This is an approximate expression, which actually does not coincide with any of the sets of values in ref.~\cite{DelNobile:2013sia} (for updated values of the coefficients, see ref.~\cite{Ellis:2018dmb}). We use it for the sake of comparison with the results from refs.~\cite{Cerdeno:2016sfi, CONUS:2021dwh}. Current uncertainties on the low-energy coefficients and masses lead to variations of 20\%-30\% in eq.~\eqref{eq:qscal}; we have numerically checked that this leads to a $\lesssim 10\%$ effect on our constraints.}

Scalar exchange also gives a contribution to the cross section for neutrino-electron elastic scattering as~\cite{Cerdeno:2016sfi}
\begin{equation}
\label{eq:csscale}	
\frac{\dd \sigma^{\rm ES}_\phi}{\dd T_e} = Z_{\rm eff}^{\rm X}(T_e) \, \frac{ g_\phi^4 \, (q^\nu_\phi)^2 \, (q^e_\phi)^2}{4 \, \pi} \, \frac{m_e^2 \, T_e}{E_\nu^2 \, (2 \, m_e \, T_e + M_\phi^2)^2} ~.
\end{equation}

In order to quantify the contribution of the scattering off electrons, in section~\ref{sec:results} we study the constraints on two specific models: a first one in which the scalar couples universally to all relevant fermions (hereafter referred to as \emph{universal}), and another one in which it only couples to leptons (dubbed \emph{leptonic scalar} model). For convenience, table~\ref{tab:charges} summarizes the explicit values of charges considered for the different SM fermions.

\begin{table}
	\centering
	\begin{tabular}{|l|cccc|}
		\hline 
		& & & & \\[-10pt]
		\qquad \qquad \; Model & \; $q^q$ \; & \; $q^e$ \; & \; $q^{\nu_e}$ \; & \; $q^{\nu_\mu}$ \; \\[2pt] \hline 
		& & & & \\[-10pt]
		Universal scalar or vector \; & 1 & 1 & 1 & 1 \\[7pt] 
		Leptonic ($\ell$) scalar & 0 & 1 & 1 & 1 \\[7pt]
		$L_e$ vector & 0 & 1 & 1 & 0 \\[7pt] 
		$B-L$ vector & $\frac{1}{3}$ & -1 & -1 & -1 \\[3pt] \hline
	\end{tabular}
	\caption{Charges for the scalar and vector mediator models considered in this work.}
	\label{tab:charges}
\end{table} 

\subsection{Light vector mediators}
\label{sec:fravec}

The Lagrangian of a simplified model of interaction of a neutral vector $Z'$ with the fermions of the first generation is given by
\begin{equation}  
\label{eq:lagvec} 	
{\cal L}_{Z'} = g_{Z'}\;Z'_\mu\left( q_{Z'}^u\, \bar{u}\gamma^\mu u+q_{Z'}^d\, \bar{d}\gamma^\mu d + q_{Z'}^e\, \bar{e} \gamma^\mu e + q_{Z'}^\nu\, \bar{\nu}_{L}\gamma^\mu \nu_{L} \right) + \frac{1}{2} M^2_{Z'} {Z'}^\mu Z'_\mu ~,
\end{equation}
where $q_{Z'}^i$ indicates the charges of each fermion ($i = \{\nu,\, e,\, u,\, d\}$) under the new interaction.

Unlike the magnetic-moment and scalar interactions, a neutral vector interaction interferes with the SM. The additional contribution to the neutrino-nucleus scattering cross section reads~\cite{Cerdeno:2016sfi}
\begin{equation}
\label{eq:csvecn}
\Delta \frac{\dd \sigma^{\rm CE\nu NS}_{Z'}}{\dd E_R} = \frac{g^2_{Z'} \, m_A}{2 \, \pi} \left[\frac{g^2_{Z'} \, (q_{Z'}^\nu)^2 \, \mathcal{Q}^2_{Z'}}{\left(2 \, m_A \, E_R + M_{Z'}^2\right)^2}
 -\frac{2 \, \sqrt{2} \, G_F \, q_{Z'}^\nu \,\mathcal{Q}_{Z'}\,\mathcal{Q}_{W}}{\left(2 \, m_A \, E_R + M_{Z'}^2\right)} \right] \left(1-\frac{m_A\,E_R}{2\,E^2_\nu}\right) |F(q)|^2 ~, 
\end{equation}
where $\mathcal{Q}_{Z'}$ is the weak charge of the nucleus for the light vector interaction. Vector current conservation implies that only valence quarks contribute by simply summing up their charges, so for universal couplings ($q_{Z'}^u = q_{Z'}^d\equiv q_{Z'}^q$), $\mathcal{Q}_{Z'} = 3 \, q_{Z'}^q \, (Z+N)$ (see, e.g., ref.~\cite{DelNobile:2013sia}). As in the case of scalar mediators, we use the same nuclear form factor as in the SM.

The corresponding contribution to the cross section for ES reads~\cite{Cerdeno:2016sfi}
\begin{equation}
\label{eq:csvece}	
\Delta \frac{\dd \sigma^{\rm ES}_{Z'}}{\dd T_e} =  Z_{\rm eff}^{\rm X}(T_e) \, \frac{g^2_{Z'} \, m_e}{2 \, \pi} \left[ \frac{{g^2_{Z'}}\, (q_{Z'}^\nu)^2 \, (q^e_{Z'})^2}{\left(2 \, m_e \, T_e + M_{Z'}^2\right)^2} \,+\,\frac{2 \, \sqrt{2} \, G_F \, q_{Z'}^\nu \, q_{Z'}^e\, c_V}{\left(2 \, m_e \, T_e + M_{Z'}^2\right) } \right] ~,
\end{equation}
where $c_{V}$ is the SM effective vector coupling introduced in eq.\eqref{eq:sm-elec}, which depends on the flavor of the incident neutrino as given in table~\ref{tab:cs}.
  
As in the case with light scalar mediators, in section~\ref{sec:results} we study the constraints on three specific models in which the vector couples universally to all relevant fermions, another in which it only couples to electron flavor ($L_e$), and the anomaly-free flavor-universal model with coupling to $B-L$ (see table~\ref{tab:charges}).

\section{Data analysis}
\label{sec:fit}

In this section we describe the procedure we follow to perform the data analysis of the \dresden and COHERENT data.

\subsection{The \dresden reactor experiment}
\label{sec:dresden}

We use a sample with 96.4-day exposure of a 2.924~kg ultra-low noise p-type point contact (PPC) germanium detector (NCC-1701) to the high flux of electron antineutrinos from the Dresden-II boiling water reactor (BWR). The new data set spans the period between January 22 and May 8, 2021, when the reactor was operated at its full nominal power of 2.96~GW. In the data release~\cite{Colaresi2022suggestive}, which we follow closely, the data points and errors are provided in the form of rate per day. The errors (per day) represent a combination of statistical and signal acceptance uncertainties. With the new information obtained from the reactor operator, the improved center-to-center effective distance between the PPC crystal and the center point of the BWR core is 10.39~m. The estimate of the $\bar{\nu}_e$ flux from the reactor is $4.8 \times 10^{13}~\bar{\nu}_e/\textrm{cm}^2 \textrm{s}$ with a $\sim 2\%$ uncertainty~\cite{Huber:2011wv, Mueller:2011nm}. We describe the reactor $\bar{\nu}_e$ spectrum using the typical average values of fission fractions of the four main isotopes (making up more than 99\% of all reactor neutrinos), for commercial power reactor using low-enriched uranium: $^{235}\rm{U}$ (58\%), $^{239}\rm{Pu}$ (29\%), $^{238}\rm{U}$ (8\%) and $^{241}\rm{Pu}$ (5\%)~\cite{Qian:2018wid}. We use the combination of the tabulated spectra for $E_{\bar{\nu}_e} < 2$~MeV from ref.~\cite{Vogel:1989iv} and for $2~\textrm{MeV} \le E_{\bar{\nu}_e} \le 8$~MeV from ref.~\cite{Mueller:2011nm}. We set the flux to zero beyond the maximum energy that is tabulated.

The background model consists of four components~\cite{Colaresi2022suggestive}, although only two of them really contribute to the SM signal region. The other two allow constraining these components. The dominant source in the signal region is the elastic scattering of epithermal neutrons, which is modeled by a free exponential plus a free constant term, $R_{\rm epith} + A_{\rm epith} \, e^{- E_{\rm rec}/E_{\rm epith}}$, where $E_{\rm rec}$ is the reconstructed ionization energy. The other three components of the background consist of electron capture (EC) peaks in $^{71}$Ge, which are all described as Gaussian probability density functions (PDF) with three free parameters: the amplitude $A_{\rm shell}$, the centroid $E_{\rm shell}$, and the standard deviation $\sigma_{\rm shell}$. The $M$-shell EC peak is constrained by the $L_1$-shell EC peak, with parameters $\{A_{L_1}, \, E_{L_1}, \, \sigma_{L_1}\}$. Although the $L_1$-shell peak is at a nominal energy of 1.297~keV, it is also allowed to vary freely. The ratio of the amplitudes of the $L_1$-shell and $M$-shell has been experimentally determined to be $0.16\pm0.03$~\cite{SuperCDMS:2015eex}, so following the data release~\cite{Colaresi2022suggestive}, we model this ratio with a Gaussian prior of width $\sigma_{M/L_1} = 0.03$ and centered at $A_M/A_{L_1} = 0.16$, which we add to the likelihood. The centroid of the $M$-shell EC contribution is fixed at $E_{M} = 0.158$~keV~\cite{SuperCDMS:2015eex, Firestone1996} and the standard deviation is set to be equal to the electronic noise $\sigma_{M} = \sigma_n$, which is 68.5~eV during the 96.4 days of reactor operation (Rx-ON) and 65.25~eV during the 25 days of reactor refueling outage (Rx-OFF). Finally, the last contribution comes from the $L_2$-shell EC, with the amplitude fixed by $A_{L_2}/A_{L_1} = 0.008$, the centroid at $E_{L_2} = 1.142$~keV and the standard deviation $\sigma_{L_2} = \sigma_{L_1}$. Explicitly, the background model, in terms of the reconstructed ionization energy, $E_{\rm rec}$, is described by a differential rate
\begin{equation}
\label{eq:bkg}
\frac{\dd R_{\rm bkg}(\boldsymbol{\beta})}{\dd E_{\rm rec}} = R_{\rm epith} + A_{\rm epith} \, e^{- E_{\rm rec}/E_{\rm epith}} + \sum_{i = L_1, L_2, M} \frac{A_i}{\sqrt{2 \, \pi} \, \sigma_i} \, e^{- \frac{\left(E_{\rm rec} - E_i\right)^2}{2 \, \sigma_i^2}} ~,
\end{equation}
in the reconstructed energy region of interest (ROI) $E_{\rm rec} = [0.2, 1.5]~{\rm keV}$. The number of free background parameters to be fitted in every analysis is seven, which are represented by the vector $\boldsymbol{\beta} = \left\{ R_{\rm epith}, A_{\rm epith}, E_{\rm epith}, A_{L_1}, E_{L_1}, \sigma_{L_1}, \beta_{M/L_1} \right\}$. The parameter $\beta_{M/L_1}$, defined as $A_M = \beta_{M/L_1} \, A_{L_1}$, is added to the likelihood with a Gaussian prior, as described below.

The CE$\nu$NS signal rate from the reactor $\bar{\nu}_e$ flux is given by
\begin{equation}
\label{eq:rate}	
\frac{\dd R_{\rm sig}^{\textrm{CE}\nu\textrm{NS}}}{\dd E_{\rm rec}} = N_{\rm T} \,
\int_{E_{\nu,{\rm min}}}^\infty \dd E_{\bar{\nu}_e} \int_{E_{R, \rm
    min}}^{E_{R, \rm max}} \dd E_R \, \frac{\dd \Phi_{\bar{\nu}_e}}{\dd
  E_{\bar{\nu}_e}} \, \frac{\dd \sigma^{\rm CE\nu NS}}{\dd E_R} \, {\cal R}(E_{\rm
  rec}, E_I=Q \, E_R ; \sigma_I) ~,
\end{equation}
where $N_T = 2.43 \times 10^{25}$ is the number of germanium nuclei and $\dd \Phi_{\bar{\nu}_e}/\dd E_{\bar{\nu}_e}$ is the reactor electron antineutrino spectrum. Here $E_{R, \rm min}$ is the minimum nuclear recoil energy which corresponds to a minimum average ionization energy, $E_{I, \rm min} \simeq 2.98$~eV, required to produce a hole-pair in germanium (at 77~K)~\cite{Antman1966272} (see ref.~\cite{Wei:2016xbw} for a compilation of existing data at other temperatures). The ionization energy is defined as $E_I = Q(E_R) \, E_R$, with $Q(E_R)$ being the quenching factor, which describes the reduction in ionization yield of a nuclear recoil when compared to an electron recoil of same energy. We consider two models from the data release, which are not in tension with CONUS data~\cite{CONUS:2020skt}, and are denoted by `Fef' (using iron-filtered monochromatic neutrons) and `YBe' (based on photoneutron source measurements)~\cite{Collar:2021fcl}. The spread between these two models approximately represents the uncertainty on this parameter. Note that the cross sections in eqs.~\eqref{eq:mag-nuc} and~\eqref{eq:mag-elec} diverge as the recoil energy goes to zero. Consequently, the contribution arising from the scattering due to the neutrino magnetic moment (and similarly for models with very light mediators) is larger in the lower energy bins, a contribution that can be divergently large under the assumption that asymptotically low ionization energies could trigger the detector. This unphysical behavior is cut-off by the physical requirement of the average energy required to produce an electron-hole pair. Thus, in what follows, we impose $E_{I,\rm min} = 3$~eV when evaluating the expected rate of events.

The energy resolution function, ${\cal R}(E_{\rm rec}, E_I ; \sigma_I)$, is described as a truncated Gaussian ($E_{\rm rec} > 0$),
\begin{equation}
\label{eq:resolutionD}	
{\cal R}(E_{\rm rec}, E_I ; \sigma_I) = \left( \frac{2}{1 + \textrm{Erf}\left(\frac{E_I}{\sqrt{2} \, \sigma_I} \right)} \right) \, \frac{1}{\sqrt{2 \, \pi} \, \sigma_I} \, e^{- \frac{\left( E_{\rm rec} -E_I\right)^2}{2 \, \sigma_I^2}} ~, 
\end{equation}
with $\sigma_I^2 = \sigma_n^2 + E_I \, \eta \, F$, where the electronic noise is $\sigma_n = 68.5~{\rm eV}$ (Rx-ON) and 65.25~eV (Rx-OFF), $\eta$ is the average energy of electron-hole formation and $F$ is the Fano factor. For Ge, $\eta = 2.96$~eV and $F = 0.11$~\cite{Colaresi2022suggestive}. The prefactor with the error function is included to guarantee the resolution function is normalized to 1 for $E_I > 0$. Note that, although the reconstructed energy ROI is $E_{\rm rec} = [0.2, 1.5]~{\rm keV}$, events with $E_I < 0.2~{\rm keV}$ are assumed to trigger the detector and be susceptible of filtering into the ROI.

Moreover, it is important to point out that, as provided, all data points are already corrected for signal acceptance, so this must not be included in the calculation of the expected signal rate. Ideally, raw data and signal acceptance as a function of the true ionization energy, $E_I$, must be used. In this way, the signal acceptance must be included in eq.~\eqref{eq:rate}. Nevertheless, the approach used here, following the data release, has a negligible impact on signals that grow slowly at small ionization energies, as the SM or NSI cases. Yet, it could have a non-negligible effect on the event rate in models with a large neutrino magnetic moment or in the presence of light mediators (in the case of interactions with nucleons). Furthermore, the impact of using signal acceptance-corrected data also depends on the minimum ionization energy capable of triggering the detector. Note that the minimum ionization energy at which the signal acceptance is currently measured is 0.13~keV~\cite{Colaresi2022suggestive}.

With all these ingredients, the expected event rate is given by
\begin{equation}
\label{eq:totalrate}
\frac{\dd R \left(\boldsymbol{\theta}; \boldsymbol{\beta}\right) }{\dd E_{\rm rec}} = \frac{\dd R_{\rm bkg} \left(\boldsymbol{\beta}\right) }{\dd E_{\rm rec}} + \frac{\dd R_{\rm sig}^{\textrm{CE}\nu\textrm{NS}} \left(\boldsymbol{\theta}\right) }{\dd E_{\rm rec}} + \frac{\dd R_{\rm sig}^{\rm ES} \left(\boldsymbol{\theta}\right) }{\dd E_{\rm rec}} ~,
\end{equation}
where our notation explicitly indicates that the event rates include both the signal contribution from CE$\nu$NS and from ES.

In general, when considering different new physics scenarios, the signal rate depends on a set of (one or two in this work) parameters $\boldsymbol{\theta}$. We use the following $\chi^2$,
\begin{equation}
\label{eq:chi2}
\chi_{\rm D-II}^2 \left(\boldsymbol{\theta}; \boldsymbol{\beta}\right) = \sum_{i=1}^{130} \frac{\left(P_i\left(\boldsymbol{\theta}; \boldsymbol{\beta}\right) - D_i\right)^2}{\sigma_i^2} + \frac{\left(\beta_{M/L_1} - A_M/A_{L_1}\right)^2}{\sigma_{M/L_1}^2} ~,
\end{equation}
where $P_i$ refers to the prediction and $D_i$ to the measured rates (or number of events in the exposure time) in energy bin $i$ (of width 10~eV and with the center of the bin indicated in the data release), $\sigma_i$ is the standard deviation of the rate (or of the measured number of events in the exposure time) in bin $i$, which combines statistical and signal acceptance uncertainties, and $\beta_{M/L_1}$ is the ratio between the amplitudes of the $M$-shell and $L_1$-shell EC contributions to the background, with central value $A_M/A_{L_1} = 0.16$ and $\sigma_{M/L_1} = 0.03$. We do not include additional nuisance parameters. Nevertheless, we have checked that even adding a 10\% uncertainty on the normalization of the signal, the results are only affected at the percent level. Finally, in order to study the sensitivity to the new physics parameters, we profile the above likelihood,
\begin{equation}
\label{eq:profchi2}
\left(\chi^2_{\rm D-II}\right)_{\rm p} \left(\boldsymbol{\theta}\right)= \chi_{\rm D-II}^2 \left(\boldsymbol{\theta}; \widehat{\widehat{\boldsymbol{\beta}\,}} (\boldsymbol{\theta}) \right) \equiv \textrm{min}_{\boldsymbol{\beta}} \left\{ \chi_{\rm D-II}^2 \left(\boldsymbol{\theta}; \boldsymbol{\beta}\right) \right\} ~,
\end{equation}
where the profiled values of $\boldsymbol{\beta}$ that minimize $\chi^2$ for each $\boldsymbol{\theta}$ are indicated by a double hat.

\begin{figure}[t]
	\begin{center}
		\includegraphics[width=\textwidth]{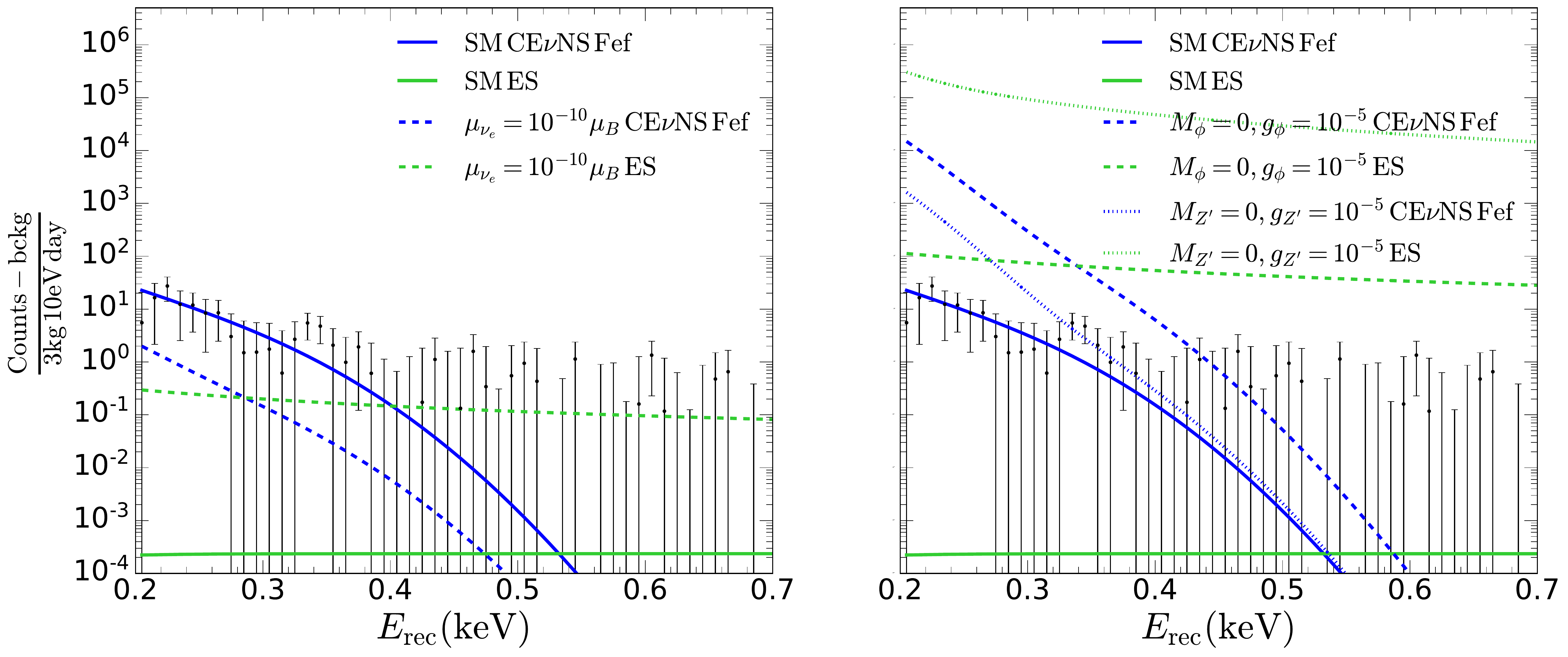} 
	\end{center} 
	\caption{\label{fig:events-dresden} \textit{Left panel:} Spectral rate of signal events from CE$\nu$NS and $\bar{\nu}_e e^-$ scattering in the SM and from the electromagnetic scattering induced by a neutrino magnetic moment $\mu_{\nu_e} = 10^{-10}\mu_B$. \textit{Right panel:} Spectral rate of signal events from CE$\nu$NS and $\bar{\nu}_e e^-$ scattering in the SM and for the contribution induced by an interaction mediated by a massless scalar boson (dashed lines) and by a massless vector boson (dotted lines), assuming universal couplings (see table~\ref{tab:charges}). For a vector mediator, we show the addition of both contributions, the purely vector boson contribution and the interference with the SM (see eq.~\eqref{eq:csvecn}). In both panels we compare the spectra to data, after subtracting the best-fit background model assuming only the SM signal. All results in both panels are depicted for the Fef model of the quenching factor (for interactions with nuclei) and assuming $E_{I,\rm min} = 3$~eV.}
\end{figure}

In figure~\ref{fig:events-dresden} we show the predicted event distributions for the SM and for the models of new physics we consider. Along with these spectra, we also include the \dresden data points, after subtracting the best-fit background model assuming only the SM signal. In the left panel of figure~\ref{fig:events-dresden} we show the predicted event distributions for the SM and the extra contribution from a non-zero magnetic moment. We see that the $\mu_{\nu_e}$-induced electromagnetic scattering off protons dominates over the corresponding scattering off electrons for $E_{\rm   rec}\lesssim 0.35$~keV. This is so because the coherent cross section off nuclei is enhanced by the factor $Z^2/Z_{\rm eff}^{\rm Ge}$, which is large enough to compensate the suppression due to the quenching factor (which makes the characteristic $1/E_R$ of the nucleus smaller than $1/T_e$ of the electron, for the same value of $E_{\rm rec}$). Also, from eqs.~\eqref{eq:csscaln},~\eqref{eq:csscale},~\eqref{eq:csvecn} and~\eqref{eq:csvece} we expect different spectra of events for interactions mediated by light scalars or by light vectors, as well as a different scaling with energy for scattering off electrons and off nucleus. This can be seen from the comparison among the different curves in the right panel of figure~\ref{fig:events-dresden}, where we have assumed universal couplings (see table~\ref{tab:charges}). The event spectra for interactions mediated by a massless scalar are very similar to the $\mu_{\nu_e}$-induced ones, as they are governed by the $1/E_R$ or $1/T_e$ dependence for CE$\nu$NS or ES, respectively. In this case, the coherent cross section off nuclei is enhanced by the factor $\mathcal{Q}_\phi^2/Z_{\rm eff}^{\rm Ge}$ with respect to that of scattering off electrons. On the other hand, the event spectra for ES mediated by a massless vector is much larger than the one for scattering off nuclei. Unlike the cases of the magnetic moment or the scalar mediator, the dominant contribution at small energies for CE$\nu$NS is a factor $\sim m_e/m_A$ smaller than for ES.

\pagebreak
\subsection{The COHERENT experiment}
\label{sec:coh}

For the COHERENT CsI analysis, we follow the same procedure as in ref.~\cite{Coloma:2019mbs}. We use the nuclear form factor from ref.~\cite{Klos:2013rwa}, we estimate the time dependence of the background directly from anti-coincidence data, and for concreteness, we use the quenching factor from ref.~\cite{Collar:2019ihs}. In the present work we also include scattering off electrons. In this case, there is no quenching factor, the target mass is set to the electron mass, and we use an effective number of electrons per nucleus, $Z_\mathrm{eff}(T_e)$, as provided in appendix~\ref{app:levels} (see also section~\ref{sec:frameworks}). The rest of the details of the analysis are the same as in ref.~\cite{Coloma:2019mbs}. Here, the minimum recoil energy considered can be safely set to zero, since the efficiency vanishes at low energies.

For the analysis of Ar data, we follow the official data release~\cite{COHERENT:2020ybo}.  In this case we do \emph{not} include scattering off electrons, as the experiment can discriminate nuclear from electron recoils~\cite{COHERENT:2020iec}. To obtain the CE$\nu$NS event spectrum, we start by computing the signal event spectrum in nuclear recoil energy, $E_R$, which is given by
\begin{equation}
\frac{\dd N}{\dd E_R} = \mathcal{N} \sum_\alpha \int_{E_{\nu,\mathrm{min}}}^{m_\mu / 2} \frac{\dd\sigma (E_\nu, E_R)}{\dd E_R} \, \frac{\dd\phi_\alpha}{\dd E_\nu} \, \dd E_\nu ~,
\end{equation}
where $E_\nu$ is the neutrino energy and the sum extends over the three components of the neutrino flux $\{\nu_e, \nu_\mu, \bar\nu_\mu\}$. The neutrino spectra $\mathrm{d}\phi_\alpha/\mathrm{d}E_\nu$ are normalized to 1 (see, e.g., eq.~(2.1) in ref.~\cite{Coloma:2019mbs} for the expressions), and all normalizations are absorbed into an overall constant $\mathcal{N}$ given by
\begin{equation}
\mathcal{N} = \frac{1 }{4\pi \ell^2} \, N_\mathrm{PoT} \, f_{\pi/p} \, N_{\mathrm{Ar}} ~,
\end{equation}
where $\ell = 27.5~\mathrm{m}$ is the distance to the detector; $N_\mathrm{PoT} = 13.77\times 10^{22}$ is the number of protons on target (PoT), corresponding to an integrated power of 6.12~GW$\cdot$hr; $f_{\pi/p} = 0.09$ is the number of pions produced per PoT; and $N_\mathrm{Ar} = m_\mathrm{det} / m_{\rm Ar}$ is the number of nuclei in the detector, with $m_\mathrm{det} = 24.4~\mathrm{kg}$ the detector mass and $m_{\rm Ar}$ the $^{40}$Ar mass.

However, the experimental collaboration does not bin their data in nuclear recoil energy (keV$_{\rm nr}$), but in electron-equivalent recoil energy instead (keV$_{\rm ee}$). In addition, we have to account for the detection efficiency, $\epsilon$, and the energy resolution. Introducing these effects, the expected event rate in each bin $i$ (of width $\Delta E_{\rm rec}$) is computed as:
\begin{equation}
\label{eq:Ni}
N_i = \int_{E_{\rm rec, i} - \Delta E_{\rm rec}/2}^{E_{\rm rec, i} + \Delta E_{\rm rec}/2} \dd E_{\rm rec} \, \, \epsilon(E_{\rm rec})\int_{E_{R, \mathrm{min}}}^\infty \dd E_R \, \frac{\dd N}{\dd E_R} \, \mathcal{R}(E_{\rm rec}, E_I; \sigma_I) ~,
\end{equation}
where $E_I$ stands for the true electron-equivalent recoil energy and $E_{\rm rec}$ is the reconstructed electron-equivalent recoil energy (that is, after energy smearing effects). The function $\mathcal{R}$ accounts for the energy resolution of the detector:
\begin{equation}
\label{eq:resolutionC}
\mathcal{R}(E_{\rm rec}, E_I; \sigma_I) = \frac{1}{\sqrt{2 \, \pi} \, \sigma_I} \, e^{\frac{-\left(E_{\rm rec} - E_I\right)^2}{2 \, \sigma_I^2}} ~,
\end{equation}
with a width $\sigma_I \equiv \sigma(E_I) = 0.58~\mathrm{keV} \, \sqrt{E_I/\mathrm{keV}}$, as prescribed in the data release. As the energies in the ROI are much larger than the standard deviation, this definition, unlike eq.~\eqref{eq:resolutionD}, does not require the extra factor to guarantee it is correctly normalized to 1. Also, note that in eq.~\eqref{eq:Ni} the detection efficiency $\epsilon$ is obtained post-triggering and it is a function of the reconstructed energy (and not of the true energy).\footnote{Furthermore, contrary to the indications provided in the data release, in our analysis we include this efficiency as a function of reconstructed energy \emph{before binning the event distribution}, as otherwise we do not find good agreement with the results of the collaboration.}

We set the minimum nuclear recoil energy to $E_{I, \mathrm{min}} = 19.5$~eV, the average energy to produce a scintillation photon in Ar~\cite{Creus:2013sau}. As indicated above, the relation between $E_I$ and $E_R$ is given by the quenching factor, $E_I = Q(E_R) \, E_R$, which is described as $Q(E_R) = a + b \, E_R$, with $a = 0.246$ and $b = 0.00078$~keV$^{-1}$, as given in the data release.

Following the procedure above, we obtain a nominal prediction of $135.3$ signal events. This is slightly higher than the rates predicted by the collaboration, but well within their reported error bars (their nominal prediction is $128\pm 17$ events). Our spectrum also shows good agreement with the official one.

To further reduce backgrounds, the analysis includes information not only on recoil energy, but also on timing and on the fraction of integrated amplitude within the first 90~ns (this last variable is called F$_{90}$ by the collaboration). Once we compute the distribution in recoil energy as described above, we can obtain the full 3D PDF. We do so by rescaling the original PDF provided by the collaboration by the ratio of their projected PDF in $E_{\rm rec}$ to our $E_{\rm rec}$ event distribution, i.e.,
\begin{equation}
N_{i,j,k} = \mathrm{PDF}_{i,j,k} \times \frac{N_{i}}{\sum_{j,k} \mathrm{PDF}_{i,j,k}} ~,
\end{equation}
where $\mathrm{PDF}_{i,j,k}$ stands for the predicted PDF provided by the collaboration as a function of recoil energy $i$, F$_{90,j}$ and time $k$. To do this, we use their nominal prediction in absence of systematics.

Finally, we include systematic uncertainties by adding nuisance parameters as prescribed in the data release.\footnote{We have checked that the impact of the quenching factor uncertainties is negligible at the level of the uncertainties quoted in the data release. Therefore, they are not included here.} For each nuisance parameter, we add a pull term to either the signal or background prediction as
\begin{equation}
n_{i,j,k} = \bar n_{i,j,k} \, (1 + \xi \, \sigma_{i,j,k}) ~,
\end{equation}
where $\bar n$ is the predicted signal or background event rates with no systematic errors, $\xi$ is the nuisance parameter, and $\sigma_{i,j,k}$ is obtained from the data release. We also add a signal normalization uncertainty as
\begin{equation}
s_{i,j,k} = \bar s_{i,j,k} \, (1 + \xi_\mathrm{norm} \, \sigma_\mathrm{norm}) ~,
\end{equation}
where we set $\sigma_\mathrm{norm}=0.1$ according to the data release (this corresponds to the ``neutrino flux'' uncertainty listed in table~1 in ref.~\cite{COHERENT:2020iec}).

We use a Poissonian $\chi^2$ for statistical uncertainties,
\begin{equation}
\left(\chi^2_{\rm COH}\right)_\mathrm{stat} = \sum_{i,j,k} 2 \left(P_{i,j,k} - D_{i,j,k} + D_{i,j,k} \ln \frac{D_{i,j,k}}{P_{i,j,k}} \right) ~,
\end{equation}
where $D$ stands for the data and $P$ for the prediction (including the effect of the nuisance parameters). A pull term is then added for each nuisance parameter $\xi_r$, as well as for the normalization of the background components,
\begin{equation}
\chi_{\rm COH}^2\left({\boldsymbol \xi}\right) =
\left(\chi^2_{\rm COH}\right)_\mathrm{stat} + \sum_r \xi_r^2 + 
\left(\frac{n_{\rm pr} - \bar n_{\rm pr}}{\sigma_{\rm pr}}\right)^2 +
\left(\frac{n_{\rm del} - \bar n_{\rm del}}{\sigma_{\rm del}}\right)^2
+ \left(\frac{n_{\rm ss} - \bar n_{\rm ss}}{\sigma_{\rm ss}}\right)^2 ~,
\end{equation}
where $\bar n$ and $\sigma$ are the central values and the uncertainties at $1\sigma$ provided in the data release for the different background components. The final $\left(\chi^2_{\rm COH}\right)_\mathrm{p} $ is obtained after minimization over all nuisance parameters and the normalization of the three background components.

\section{Results}
\label{sec:results}

In this section we present the results of the analysis of the \dresden reactor experiment data and its combination with COHERENT CsI and Ar data, for the BSM frameworks presented in section~\ref{sec:frameworks}.

\subsection{Bounds on non-standard neutrino interactions}
\label{sec:resnsi}

\begin{figure}[t]
	\begin{center}
		\includegraphics[width=\textwidth]{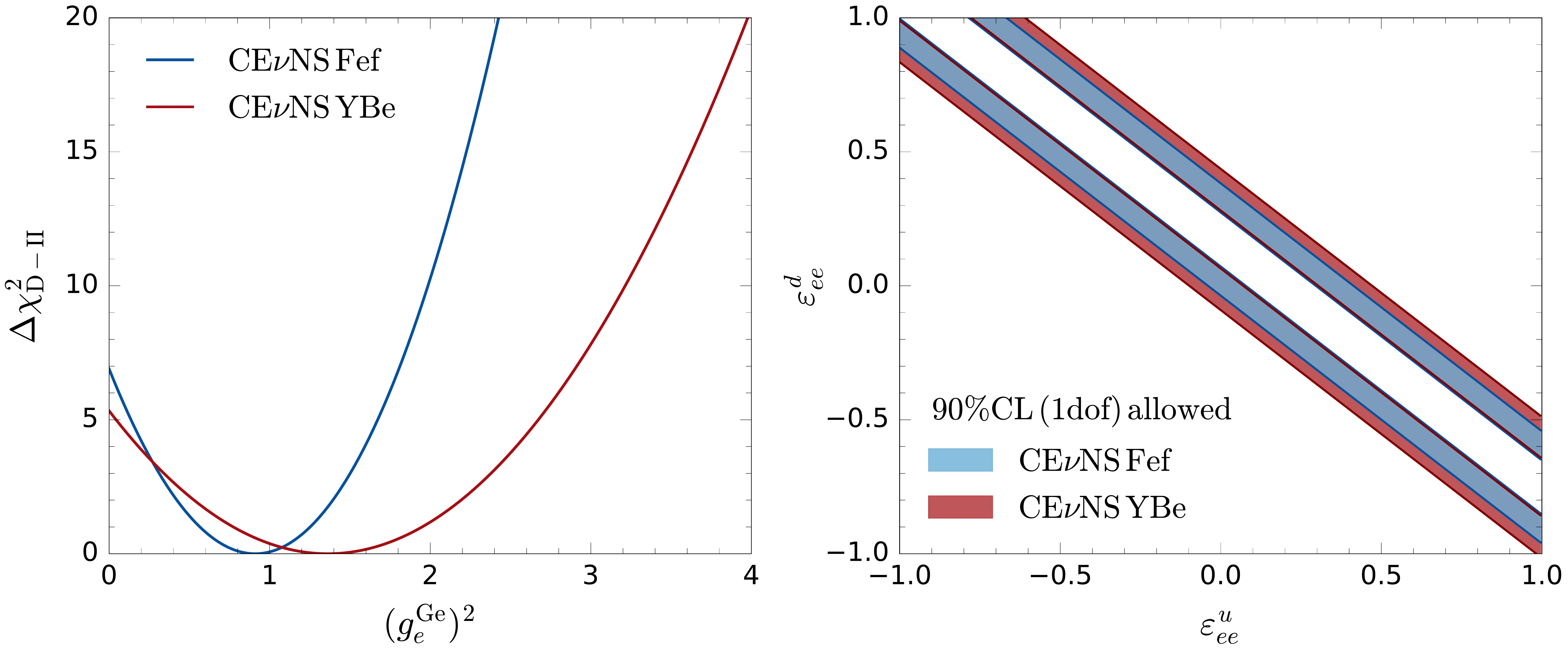}
	\end{center}
	\caption{\textit{Left panel:} Dependence of the profiled $\Delta\chi^2_{\rm D-II}$ on the combination of NSI coefficients relevant to CE$\nu$NS of $\bar\nu_e$ scattering off the Ge detector of the \dresden experiment. \textit{Right panel:} Allowed regions at 90\% CL (1 dof, two-sided, $\Delta\chi^2 = 2.71$) in the $(\Eps_{ee}^u, \Eps_{ee}^d)$ plane. In both panels, the results are shown for two quenching factors, denoted by Fef (in blue), using iron-filtered monochromatic neutrons, and YBe (in red), based on photoneutron source measurements~\cite{Collar:2021fcl}. Although not obvious, note that the allowed regions for the two quenching factors have one common side, so the Fef constraints are more stringent.}
	\label{fig:chi2_nsi}
\end{figure}

As described in section~\ref{sec:fraNSI}, in presence of NSI the \dresden reactor experiment is sensitive to a unique combination of the $\Eps$ coefficients, $\mathcal{Q}^{\rm Ge}_e(\boldsymbol\Eps)$, defined in eq.~\eqref{eq:Qalpha-nsi-X}, where we have introduced an explicit superindex indicating the nucleus it refers to. We show in the left panel of figure~\ref{fig:chi2_nsi} the dependence of $\Delta\chi^2_{\rm   D-II}$ on this effective NSI combination. In constructing this $\Delta\chi^2_{\rm D-II}$, we have profiled over the background model parameters for each value of $\mathcal{Q}^{\rm Ge}_e(\boldsymbol{\Eps})^2$ and we have neglected small effects of any additional systematic nuisance parameter. As mentioned above, the signal acceptance uncertainties are included in the data provided by the experiment.

From figure~\ref{fig:chi2_nsi}, we read the following ranges allowed at 90\% confidence level (CL) (1 dof)
\begin{equation}
\label{eq:geNSI}	
\left(g_e^{\rm Ge}\right)^2 \equiv \left(\frac{\mathcal{Q}^{\rm Ge}_e(\boldsymbol\Eps)}{\mathcal{Q}^{\rm Ge}_e(0)}\right)^2 = 0.91 \pm 0.56\;\; (1.36 \pm 0.97) ~,
\end{equation}
derived with the Fef (YBe) quenching factor for germanium. Notice that $g_e^{\rm Ge} = 1$ corresponds to the SM prediction and $g_e^{\rm Ge} = 0$ to no signal, so this implies that the absence of CE$\nu$NS in the \dresden data is disfavored at 2.6$\sigma$ (2.3$\sigma$) for the analysis with the Fef (YBe) quenching factor.

In the right panel of figure~\ref{fig:chi2_nsi}, we show the corresponding 90\% CL allowed regions for the flavor-diagonal NSI coefficients $\Eps_{ee}^u$ and $\Eps_{ee}^d$ assuming vanishing non-diagonal NSI coefficients. The shape of these bands can be understood directly from the expression of the weak charge of the nucleus in eq.~\eqref{eq:Qalpha-nsi}. They are defined as the two regions around the points that satisfy
\begin{equation}
 \left[\mathcal{Q}_{W} + (2 \, Z + N) \, \Eps_{ee}^u + (2 \, N + Z) \, \Eps_{ee}^d\right]^2 = {\rm constant} ~,
\end{equation}
which follow lines in the $(\Eps_{ee}^u, \Eps_{ee}^d)$ plane with the slope given by $-(2Z+N)/(2N+Z)$. As seen in the figure, the analysis employing the Fef quenching factor results in slightly stronger constraints.  In most scenarios we find that the analysis using the Fef quenching factor reproduces slightly better the SM predictions and therefore leaves less room for new physics. Exception, as we will see, is the case for some models with light vector mediators for which a local non-standard minima appears in the analysis with the YBe quenching factor, which results in slightly stronger constraints (see section~\ref{sec:resvec}).

\begin{figure}[t]
	\begin{center}
		\includegraphics[width=\textwidth]{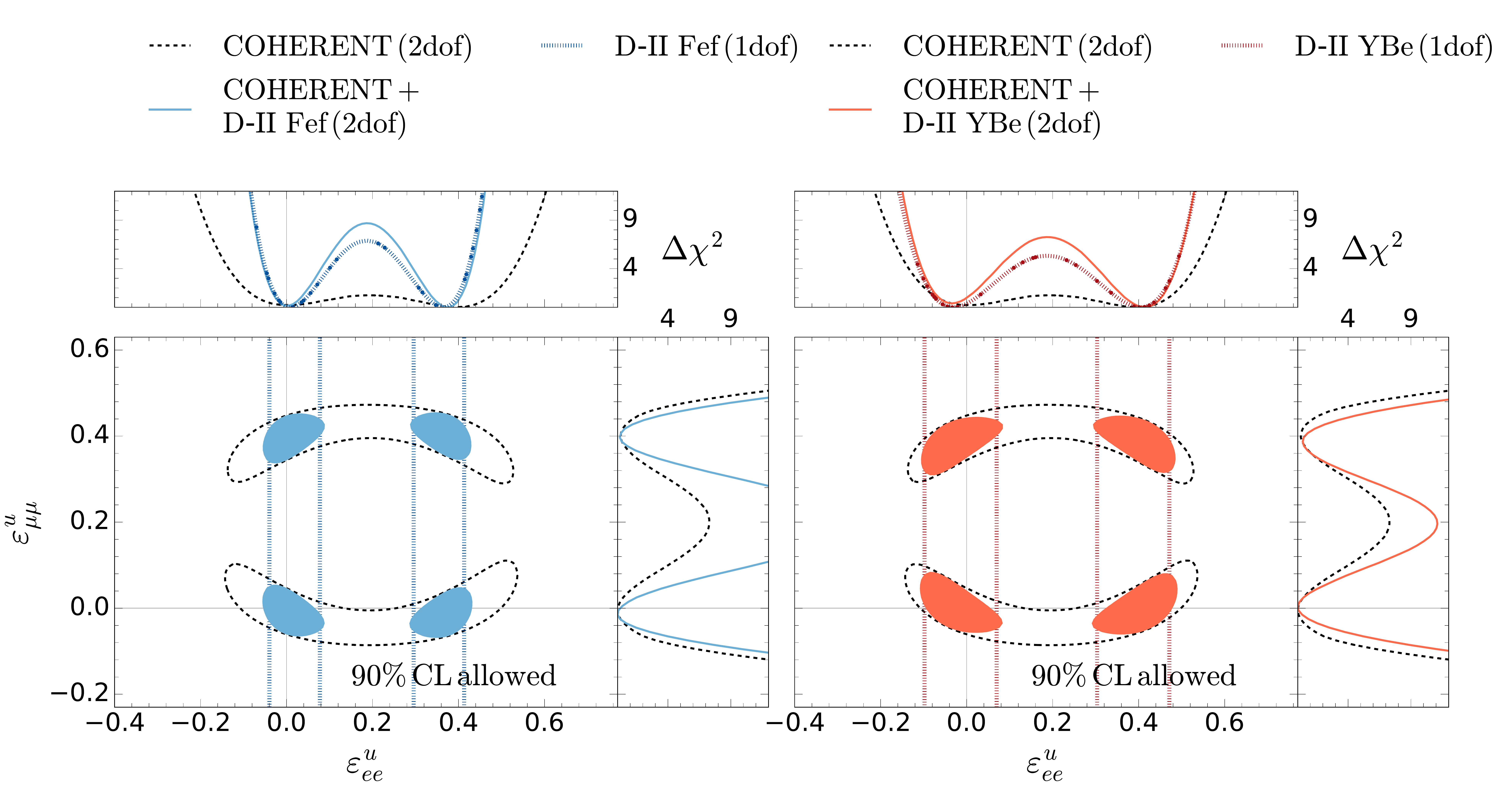}
	\end{center}
	\caption{90\% CL allowed regions on flavor diagonal NSI with up-quarks (for zero values of all other NSI coefficients) from the analysis of COHERENT CsI and Ar data, the \dresden reactor data -- with Fef (YBe) quenching factor in left (right) panel --, and their combination. Note that, in the two-dimensional panels, the results for the \dresden reactor experiment are obtained for 1 dof ($\Delta\chi^2 = 2.71$), while the rest of the regions are obtained for 2 dof ($\Delta\chi^2 = 4.61$).}
	\label{fig:nsireg}
\end{figure}

CE$\nu$NS at the COHERENT experiment is sensitive to interactions of electron and muon neutrinos and hence, it provides information on the corresponding effective combinations $\mathcal{Q}^{\rm CsI}_e(\boldsymbol\Eps)$, $\mathcal{Q}^{\rm CsI}_\mu(\boldsymbol\Eps)$, $\mathcal{Q}^{\rm Ar}_e(\boldsymbol\Eps)$, $\mathcal{Q}^{\rm Ar}_\mu(\boldsymbol\Eps)$. Generically, because both $\nu_e$ and $\nu_\mu$ are present in the beam, degeneracies between NSI parameters corresponding to $e$ and $\mu$ flavors appear. This is illustrated in figure~\ref{fig:nsireg}, where we show the allowed regions obtained from our combined analysis of CE$\nu$NS at COHERENT with both CsI and Ar targets for flavor-diagonal NSIs with up-quarks only (the results for NSI with only down-quarks are similar). The shape of these regions for a given nucleus leads to allowed regions defined by a band around the points that approximately obey the equation of an ellipse in the $(\Eps_{ee}^u, \Eps_{\mu\mu}^u)$ plane,
\begin{equation}
\label{eq:cohreg}	
\left[ \mathcal{Q}_{W} + \left( 2 \, Z + N \right) \, 
\Eps_{ee}^u\right]^2 + 2 \, \left[ \mathcal{Q}_{W} + \left( 2 \, Z + N \right) \, \Eps_{\mu\mu}^u\right]^2 = {\rm constant} ~.
\end{equation}
Since $\mathcal{Q}_{W}$ depends on the target nucleus, the ellipse obtained is different for different detector materials and, therefore, it may be broken by adding information on different nuclei, provided they have a different ratio of protons to neutrons (for earlier discussions on this, see, e.g., refs.~\cite{Scholberg:2005qs, Barranco:2005yy, Coloma:2017egw, Baxter:2019mcx}). Most importantly, the use of timing information at COHERENT translates into a partial discrimination between the weak charges for the different flavors, thanks to the distinct composition of the prompt ($\nu_\mu$) and delayed ($\bar\nu_\mu$ and $\nu_e$) neutrino flux. This leads to a partial breaking of this degeneracy which, however, is not complete (see, e.g., ref.~\cite{Coloma:2019mbs} for a more detailed explanation of this effect). This is what explains the results shown in figure~\ref{fig:nsireg}. The regions allowed by COHERENT data present a double-wedge shape due to the partial breaking of the degeneracy between those two parameters after the inclusion of the energy and, in particular, the timing information.\footnote{For concreteness, the COHERENT analysis shown in this plot was performed using the quenching factor from the Chicago group~\cite{Collar:2019ihs}. See ref.~\cite{Coloma:2019mbs} for a discussion about the small variation of the results obtained with other quenching factors.} But yet, a continuous wide range of values of $\Eps^u_{ee}$ remains allowed.

For the considered flavor-diagonal NSI with up-quarks only, CE$\nu$NS at the \dresden reactor experiment provides information on $\Eps^u_{ee}$ only, so the allowed regions correspond to the vertical bands in the figure. Consequently, the combined analysis of COHERENT + \dresden results in a substantial reduction of the allowed values of $\Eps^u_{ee}$ (and indirectly, also on $\Eps^u_{\mu\mu}$) as seen in the figure.

We finish by commenting that the results we show correspond to the case of diagonal NSI with up-quarks only, but as mentioned above, similar results are obtained for diagonal NSI with down-quarks only. We also notice that the inclusion of flavor off-diagonal NSI couplings results in the enlargement of the regions shown in figure~\ref{fig:nsireg}. Nevertheless, within the current constraints from other experiments, in particular from neutrino oscillations~\cite{Esteban:2018ppq, Coloma:2019mbs}, similar qualitative conclusions hold. For NSI couplings to both up and down quarks, the complementarity between COHERENT and \dresden data depends on the specific assumption about the ratio of diagonal couplings to both quarks. If they are varied freely, meaningful constraints cannot be obtained, as can be understood from figure~\ref{fig:chi2_nsi}. Therefore, current data from these experiments is not enough by itself to impose meaningful constraints on the complete parameter space of NSI with quarks, if considered in full generality.

\subsection{Bounds on neutrino magnetic moments}
\label{sec:resmag}

\begin{figure}
	\centering
	\includegraphics[width=\textwidth]{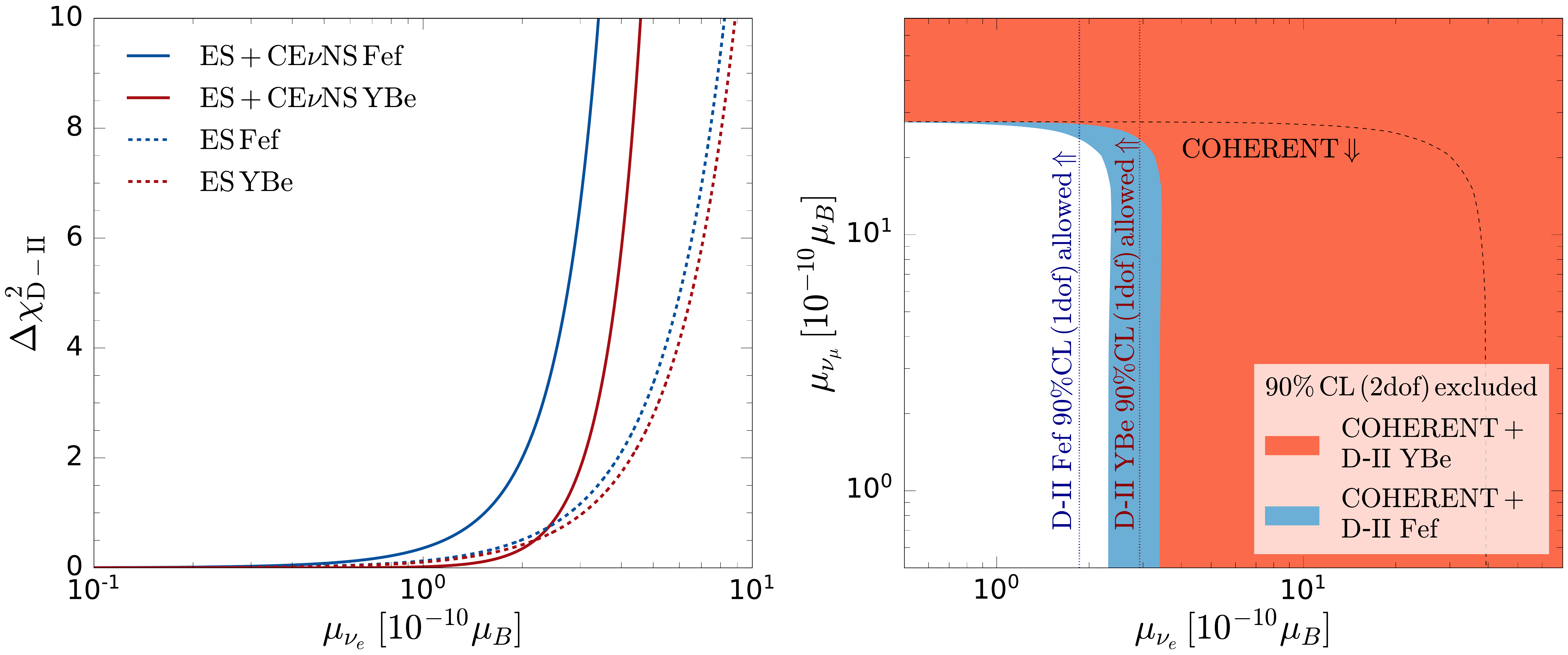}
	\caption{\textit{Left panel:} Profiled $\Delta\chi^2_{\rm D-II}$ including events induced by a magnetic moment for $\bar\nu_e$. We show results including only $\mu_{\nu_e}$-induced scattering off electrons (dashed curves) and for scattering off electrons and nucleons (solid curves). In both cases the SM CE$\nu$NS contribution is included. \textit{Right panel:} 90\% CL excluded (one-sided) regions from the combination of \dresden and COHERENT data (in color). We also indicate (with arrows) the 90\% CL (one-sided) allowed regions from \dresden data with the Fef (blue dotted line) and YBe (red dotted line)  quenching factor, and from the combined analysis of COHERENT CsI and Ar data (black dashed line). Notice that the vertical lines for the \dresden reactor experiment are defined one-sided for 1 dof $(\Delta\chi^2 = 1.64$), while the rest of the regions are defined for 2 dof ($\Delta\chi^2 = 3.22$).}
	\label{fig:chi2mag}
\end{figure}

The results of our analysis of data from the \dresden reactor experiment, including the contribution induced by a neutrino magnetic moment, are shown in figure~\ref{fig:chi2mag} where, on the left panel, we plot the one-dimensional $\Delta\chi^2_{\rm D-II}$ as a function $\mu_{\nu_e}$ after profiling over all background parameters ${\boldsymbol \beta}$. As previously, we have neglected small effects of any additional systematic nuisance parameter, while the signal acceptance uncertainties are included in the data provided by the experiment. For comparison, in the left panel we also show $\Delta\chi^2_{\rm D-II}$ for the case of scattering off electrons only, including the SM CE$\nu$NS contribution (dashed lines). As discussed in section~\ref{sec:dresden}, the contribution from ES induced by a neutrino magnetic moment is subdominant to that from CE$\nu$NS. One must notice, however, that a better bound on $\mu_{\nu_e}$ (and similarly for models with light scalar mediators, see below) could be attainable with a dedicated analysis aimed at optimizing the sensitivity to the signal from scattering of electrons. The signal acceptance at the lowest energies in the \dresden experiment is quite small~\cite{Colaresi2022suggestive}, which significantly reduces statistics. Furthermore, given the much flatter shape of the event spectrum for the magnetic moment contribution than from the SM, as depicted in figure~\ref{fig:events-dresden}, extending the ROI to higher energies could enhance the signal-to-noise ratio (see, e.g., ref.~\cite{Bonet:2022imz}).
	
The fit shows no evidence of a non-zero magnetic moment and therefore, the analysis results in a bound which, at 90\% CL (1 dof), reads
\begin{equation}
\label{eq:magbound}
\left|\mu_{\nu_e}\right| <
\left\{ \begin{array}{c}
 1.9\;(2.9)\times 10^{-10} \;\mu_B \quad \textrm{(one-sided~limit)}  \\[1.5ex]
 2.2\;(3.3)\times 10^{-10} \;\mu_B \quad \textrm{(two-sided~limit)} \end{array} 
\right.
~, 
\end{equation}
for the Fef (YBe) quenching factor for germanium. This is an order of magnitude weaker than the current best limit from reactor antineutrinos~\cite{Beda:2012zz, Beda:2013mta}. Notice that here we report two different limits according to different statistical criteria to derive constraints. The reason is that the experiment is in fact sensitive to $|\mu_{\nu_e}|^2$, which can only take non-negative values. Therefore, accounting for the physical boundary, it is possible to report the limit on $|\mu_{\nu_e}|$ as a {\sl one-sided} limit, which at 90\% CL for 1 dof (2 dof) corresponds to $\Delta\chi^2 = 1.64 \; (3.22)$. Conversely, if this restriction is not imposed, the result obtained is what is denoted as a {\sl two-sided} limit, which at 90\%CL for 1 dof (2 dof) corresponds to $\Delta\chi^2 = 2.71 \; (4.61)$ and results into less tight constraints. For the sake of comparison with different results in the literature, we indicate in eq.~\eqref{eq:magbound} bounds obtained with both criteria. As seen in eq.~\eqref{eq:magbound}, the difference is at the level of 10\%.

As mentioned in section~\ref{sec:dresden} the $\mu_\nu$-induced cross sections in eqs.~\eqref{eq:mag-nuc} and~\eqref{eq:mag-elec} diverge as the recoil energy goes to zero and this unphysical behavior is cut-off by the physical requirement of the average energy required to produce an electron-hole pair, $E_{I,\rm min} = 3$~eV. This raises the issue of the possible dependence of the bounds on the exact value of this minimum energy that could trigger the detector. The dependence on the cut energy, however, is approximately logarithmic and we have verified that increasing it by as much as one order of magnitude results in weakening the bounds in eq.~\eqref{eq:magbound} by less than 25\%.

CE$\nu$NS at the COHERENT experiment provides information on magnetic moments for both $\nu_e$ and $\nu_\mu$. In the right panel of figure~\ref{fig:chi2mag}, we show their 90\% CL (2 dof, one-sided) excluded values, obtained with our combined analysis of CE$\nu$NS at COHERENT with both CsI and Ar targets. Because there is no interference between the contributions from $\mu_{\nu_\mu}$ and $\mu_{\nu_e}$, the resulting allowed region is just a square with a rounded upper-right corner. The 90\% CL (1 dof, one-sided) upper bound from the \dresden reactor experiment on $\mu_{\nu_e}$, eq.~\eqref{eq:magbound}, is indicated by vertical lines. As seen in the figure, the sensitivity of COHERENT to $\mu_{\nu_e}$ is ${\cal O} (10)$ weaker. The combination of the two experiments results in the 90\% CL excluded regions (2 dof, one-sided) shown in the figure in color. The corresponding combined bounds on each neutrino magnetic moment (after profiling over the other) at 90\% CL (1 dof) are
\begin{eqnarray}	
\left|\mu_{\nu_e}\right| & < & 
\left\{
\begin{array}{c}
1.8\;(2.8)\times 10^{-10} \;\mu_B \quad \textrm{(one-sided~limit)} \\[1.5ex] 
2.2\;(3.2)\times 10^{-10} \;\mu_B \quad \textrm{(two-sided~limit)}
\end{array} \right. 
~, \nonumber \\[1.5ex] 
\left|\mu_{\nu_\mu}\right| & < & 
\left\{ 
\begin{array}{c}
2.4\;(2.4)\times 10^{-9} \;\mu_B \quad \;\, \textrm{(one-sided~limit)}  \\[1.5ex]
2.7\;(2.7)\times 10^{-9} \;\mu_B \quad \; \textrm{(two-sided~limit)}  
\end{array}\right.  
~, 
\label{eq:magcomb}
\end{eqnarray}     
for the combination of data from the COHERENT and the \dresden reactor experiments performed with the Fef (YBe) quenching factor for germanium.

\subsection{Bounds on light scalar mediator models}
\label{sec:ressca}

\begin{figure}[t]
	\begin{center}
		\includegraphics[width=0.49\textwidth]{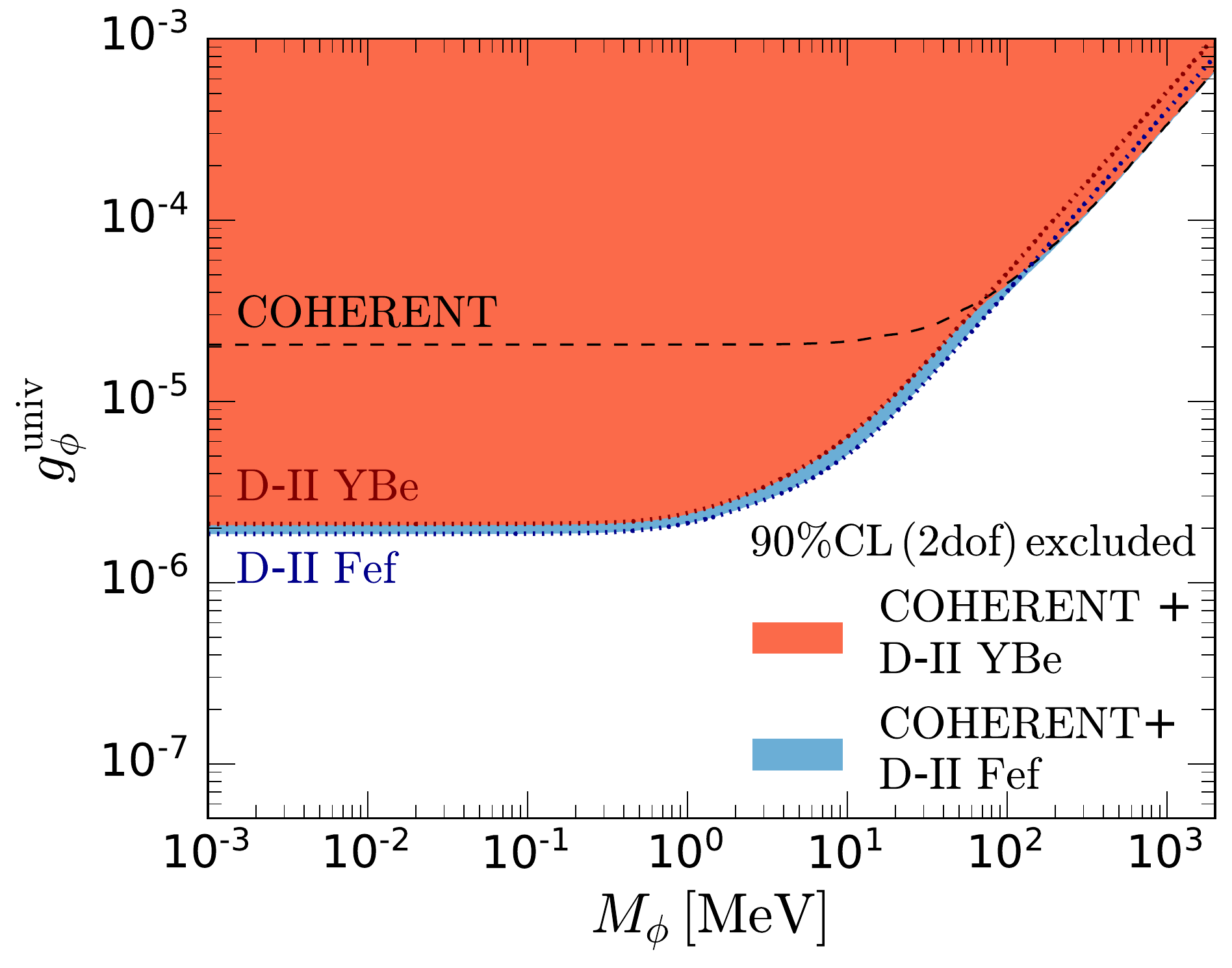}
		\includegraphics[width=0.49\textwidth]{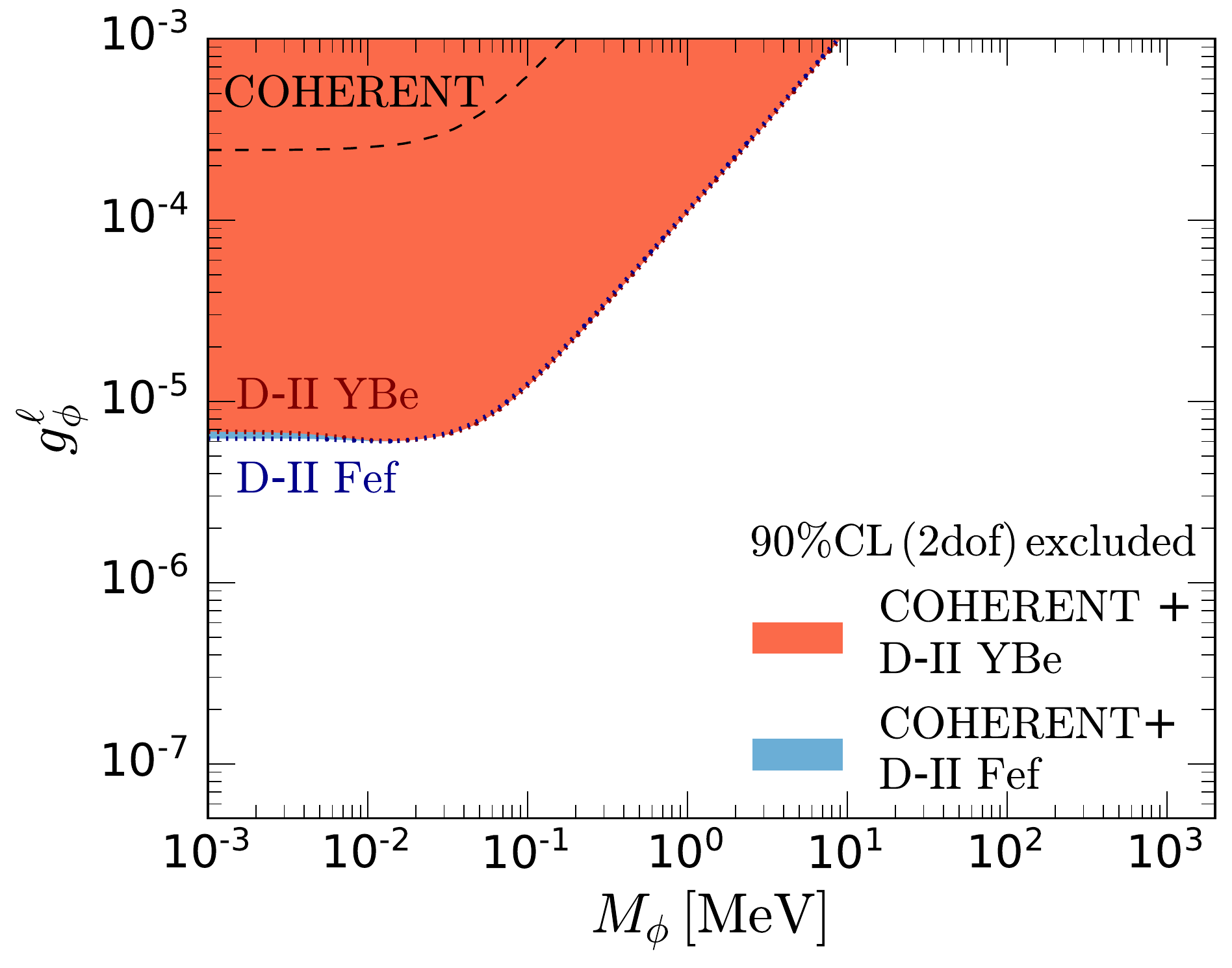}
	\end{center}
	\caption{\textit{Left panel:} 90\% CL (2 dof, two-sided, $\Delta\chi^2 = 4.61$) excluded regions for models with light scalar mediators coupled universally to all relevant fermions. \textit{Right panel:} 90\% CL (2 dof, two-sided, $\Delta\chi^2 = 4.61$) excluded regions for models with light scalar mediators coupled only to leptons. The thick blue and red dotted lines show the boundary of the region excluded by the \dresden reactor experiment with the Fef and YBe quenching factors, respectively. The black dashed line shows the boundary of region excluded from the analysis of COHERENT CsI and Ar data. The filled regions are those excluded by the combined COHERENT+\dresden analysis.}
	\label{fig:res-scalar}
\end{figure}

The results of the analysis of the data from the \dresden reactor experiment and its combination with COHERENT, including the contribution of the events induced by a new light scalar mediator, are shown in figure~\ref{fig:res-scalar}. We depict the excluded region at 90\% CL (2 dof, two-sided) for the parameter space of the scalar models $\left(\left|g^{\rm mod}_\phi\right|, M_\phi\right)$, for the two models with charges listed in table~\ref{tab:charges}. On the left panel, we show the results for a scalar coupling universally coupled to all relevant fermions, while on the right panel we focus on a scalar which only interacts with leptons.

The thick blue and red thick dotted lines in figure~\ref{fig:res-scalar} show the boundaries of the regions excluded by the \dresden reactor data with with the Fef and YBe quenching factor, respectively. These results are obtained after profiling over the background model parameters (i.e., for each value of $\left(\left|g^{\rm mod}_\phi\right|, M_\phi\right)$). For light mediator masses, the experiment has no sensitivity to the mediator mass and the limit of the regions approaches a horizontal line of constant $\left|g^{\rm mod}_\phi\right|$. For these effectively massless scalar mediators it is possible to derive an upper bound on the coupling constant regardless the mediator mass, which for the considered models, reads at 90\% CL (1 dof), 
\begin{eqnarray}
 \left|g_\phi^{\rm univ}\right| & \leq & 
 \left\{
 \begin{array}{c}
 1.6 \; (1.9) \times 10^{-6} \quad \textrm{(one-sided)} \\[1.5ex]
 1.7 \; (2.0) \times 10^{-6} \quad \textrm{(two-sided)}  
 \end{array}
 \right. ~, \nonumber \\[1.5ex]
 \left|g_\phi^{\ell}\right| & \leq & 
 \left\{
 \begin{array}{c}
 4.9 \; (5.5) \times 10^{-6} \quad \textrm{(one-sided)}  \\[1.5ex] 
 5.5 \; (6.1) \times 10^{-6} \quad \textrm{(two-sided)}    
\end{array} \right. ~,
\label{eq:scalb1}	 
\end{eqnarray}
for the Fef (YBe) quenching factor for germanium. As done above, we indicate both, one-sided and two-sided (1 dof) bounds.

Conversely, for larger mediator masses, the boundary is a diagonal, characteristic of the contact-interaction limit. In that case, the event rates depend on $(g_\phi/M_\phi)^4$, which we find to be well approximated by
\begin{eqnarray}
\frac{\left|g_\phi^{\rm univ}\right|}{M_\phi/{\rm MeV}} & \gtrsim & 3.4 \; (3.6) \times 10^{-7} ~, \quad {\rm for\;} M_\phi \gtrsim 10\;{\rm MeV} ~, \nonumber \\[1.5ex] 
\frac{\left|g_\phi^{\ell}\right|}{M_\phi/{\rm MeV}} & \gtrsim & 1.1 \; (1.1) \times 10^{-4} ~, \quad {\rm for\;} M_\phi \gtrsim 0.1\;{\rm MeV} ~,
\label{eq:scalb2}
\end{eqnarray}
for the Fef (YBe) quenching factor for germanium.

In summary, we find that for sufficiently light mediators ($M_\phi \lesssim 10^{-2}~\mathrm{MeV}$), the bound on the coupling constant is about a factor ${\cal O}(3)$ stronger if the interaction is coupled to quarks than if it couples only to leptons. This is so because scalar-mediated CE$\nu$NS always dominates over the corresponding scalar-mediated incoherent scattering off electrons for the low-energy part of the spectrum, where statistics is best (see figure~\ref{fig:events-dresden}). Equivalently, in the contact-interaction limit, the bound on $\left|g^{\rm univ}_\phi\right|/M_\phi$ is more than two orders of magnitude better than on $\left|g^{\ell}_\phi\right|/M_\phi$.

In figure~\ref{fig:res-scalar}, we also show the boundary of the regions excluded from the analysis of COHERENT CsI and Ar data (black dashed lines). For most of the parameter space we consider, the bounds imposed by the \dresden reactor experiment dominate in either model. COHERENT bounds only become competitive for the universal model for $\left|g_\phi^{\rm univ}\right| \gtrsim 5\times 10^{-5}$. Consequently, the results in eqs.~\eqref{eq:scalb1} and~\eqref{eq:scalb2} hold to good approximation for the combination of COHERENT and \dresden reactor experiments.

\subsection{Bounds on light vector mediator models}
\label{sec:resvec}

\begin{figure}[t]
\begin{center}
\includegraphics[width=0.49\textwidth]{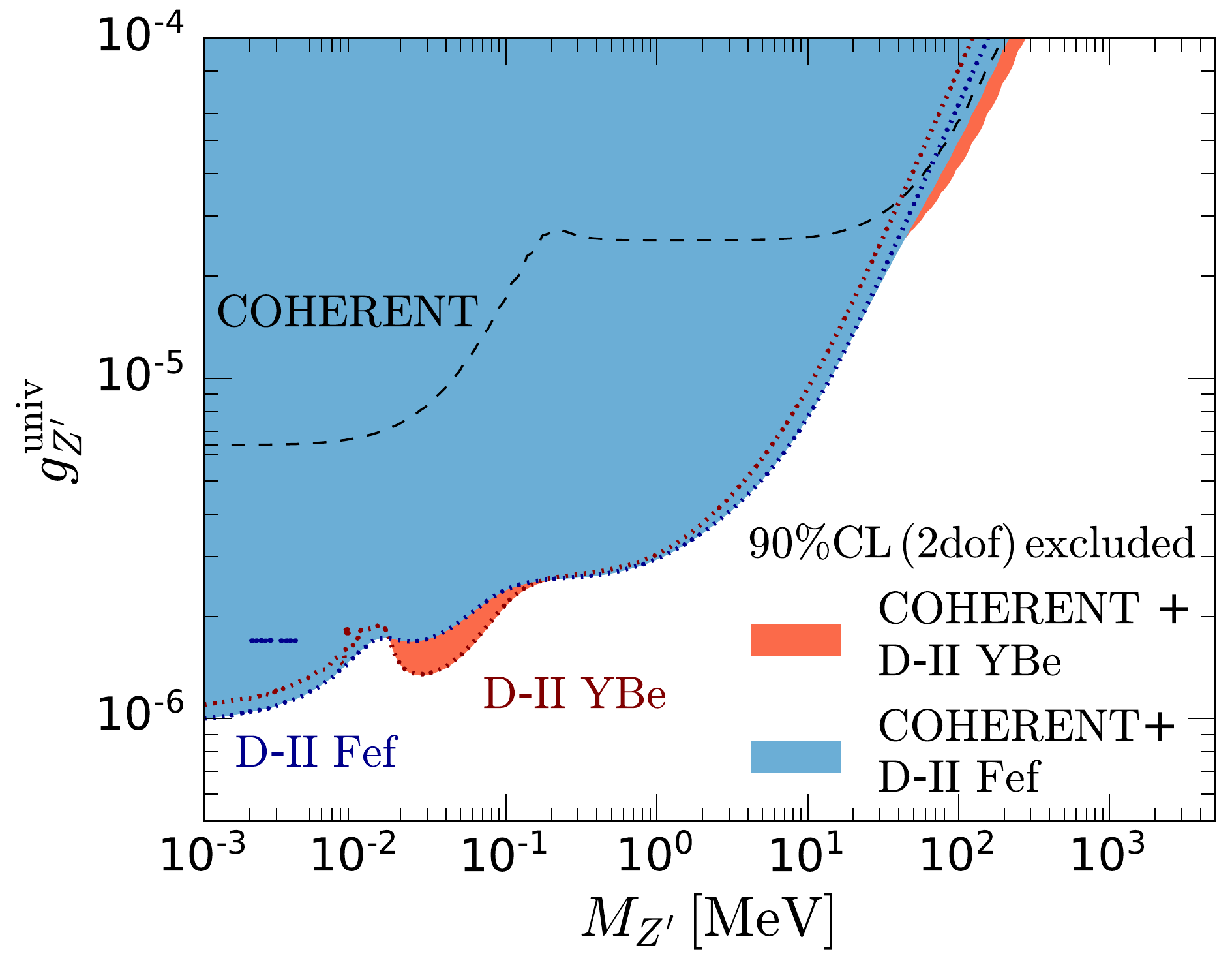}
\includegraphics[width=0.49\textwidth]{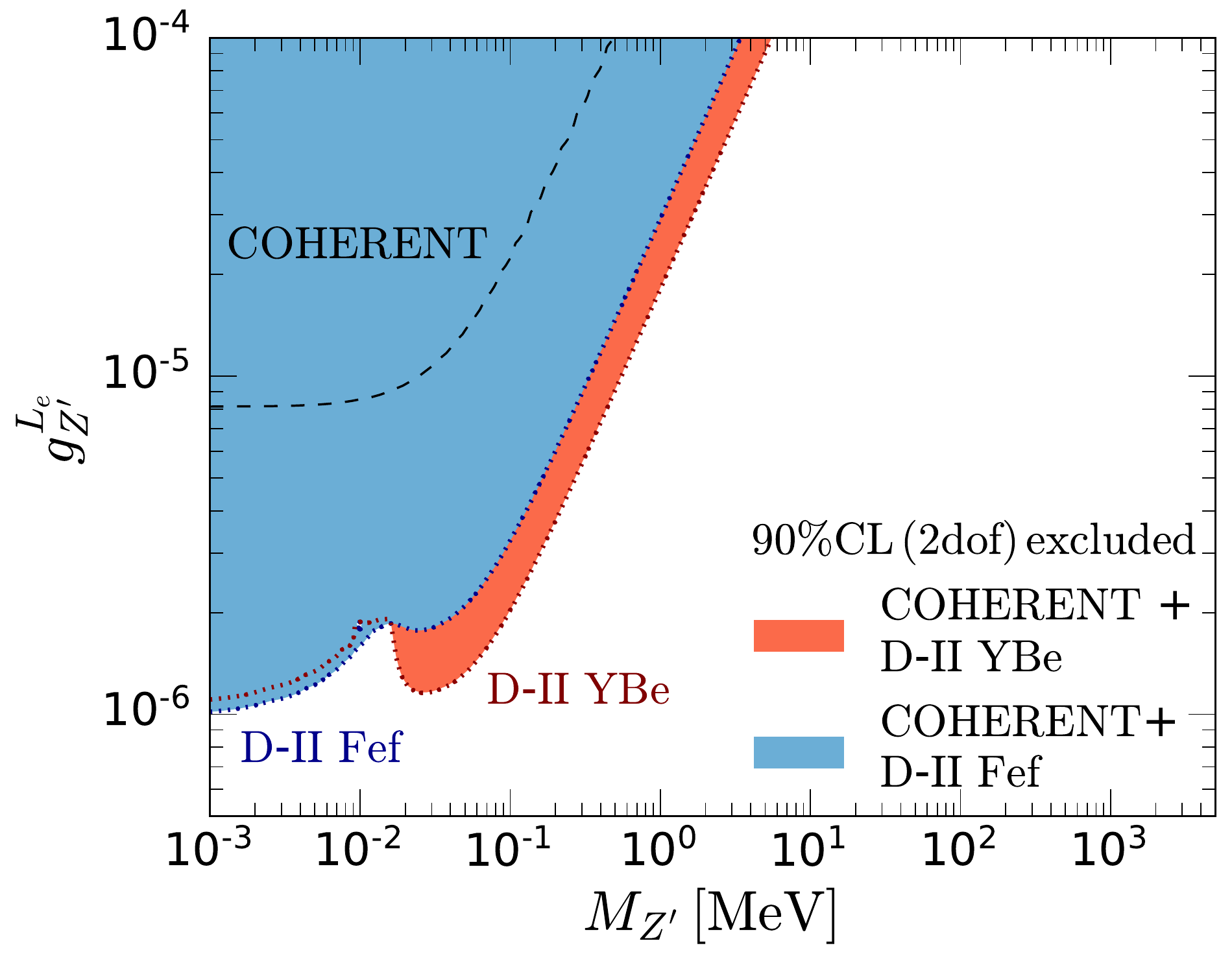}
\includegraphics[width=0.49\textwidth]{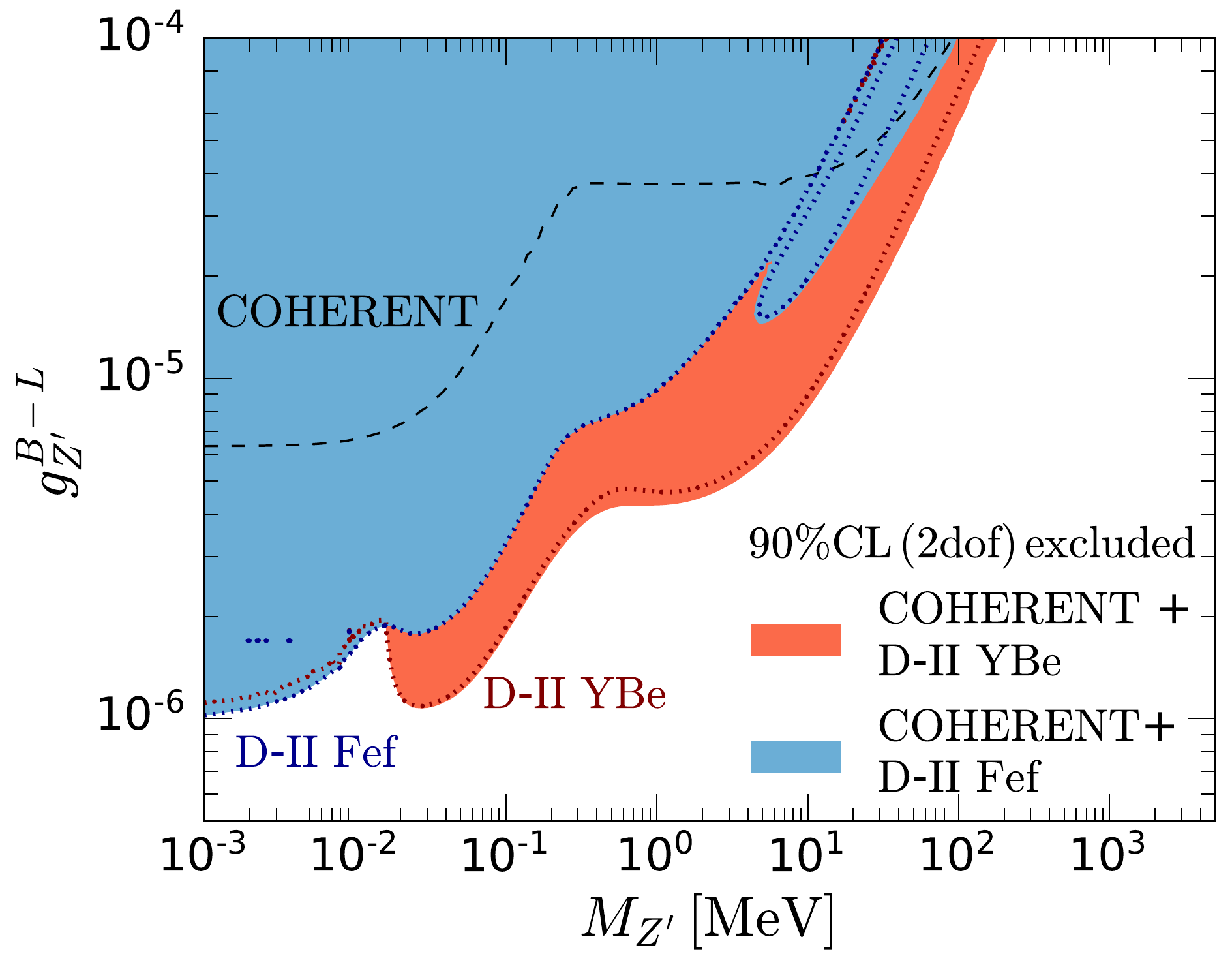}
\end{center}
\caption{\textit{Upper left panel:} 90\% CL (2 dof, two-sided, $\Delta\chi^2 = 4.61$) excluded regions for models with light vector mediators coupled universally to all relevant fermions. \textit{Upper right panel:} 90\% CL (2 dof, two-sided, $\Delta\chi^2 = 4.61$) excluded regions for models with light vector mediators coupled only to $L_e$. \textit{Lower panel:} 90\% CL (2 dof, two-sided, $\Delta\chi^2 = 4.61$) excluded regions for models with light vector mediators coupled only to $B-L$. The thick blue and red dotted lines show the boundary of the region excluded by the \dresden reactor experiment with the Fef and YBe quenching factors, respectively. The black dashed line shows the boundary of region excluded from the analysis of COHERENT CsI and Ar data. The filled regions are those excluded by the combined COHERENT+\dresden analysis.}
\label{fig:res-vector}
\end{figure}

The results of the analysis of the data from the \dresden reactor experiment and its combination with COHERENT, including the contribution of the events induced by a new light vector mediator, are shown in figure~\ref{fig:res-vector}. The figure displays the regions excluded at 90\% CL (2 dof, two-sided) corresponding to the parameter space of vector mediator models $\left(\left|g^{\rm mod}_{Z'}\right|, M_{Z'}\right)$, for the three cases with charges listed in table~\ref{tab:charges}.

As above, the thick blue and red thick dotted lines in the figures indicate the boundaries of the regions excluded at 90\% CL by the \dresden reactor data with the Fef and YBe quenching factor, respectively, after profiling over the background model parameters. The black dashed lines indicate the boundary of the regions excluded from the analysis of COHERENT CsI and Ar data and the filled regions correspond to the results of the COHERENT + \dresden combination. We see that COHERENT bounds only give some contribution for the universal and $B-L$ models for $\left\{\left|g_{Z'}^{\rm univ}\right|, \left|g_{Z'}^{B-L}\right|\right\} \gtrsim 5\times 10^{-4}$.

Again, we find that for sufficiently light mediator masses ($M_\phi \lesssim 10^{-2}~\mathrm{MeV}$) the region becomes independent of the mediator mass. For these effectively massless vector mediators, the corresponding 90\% CL (1 dof) upper bounds on the coupling constant read
\begin{eqnarray}
 \left|g_{Z'}^{\rm univ}\right| & \leq & 
 \left\{
 \begin{array}{c}
 0.87 \; (1.0) \times 10^{-6} \quad \textrm{(one-sided)} \\[1.5ex]
 0.92 \; (1.1) \times 10^{-6} \quad \textrm{(two-sided)} \nonumber  
 \end{array}
 \right. ~, \\[1.5ex]
 \left|g_{Z'}^{L_e}\right|  & \leq & 
 \left\{
 \begin{array}{c}
 0.90 \; (1.0) \times 10^{-6} \quad \textrm{(one-sided)} \\[1.5ex]
  0.95 \; (1.1) \times 10^{-6} \quad \textrm{(two-sided)} 
 \end{array}
 \right. ~, \\[1.5ex]
 \left|g_{Z'}^{B-L}\right|  & \leq & 
 \left\{
 \begin{array}{c}
 0.90 \; (1.1) \times 10^{-6} \quad \textrm{(one-sided)} \\[1.5ex]
 0.95 \; (1.1) \times 10^{-6} \quad \textrm{(two-sided)}\nonumber  
 \end{array}
 \right. ~, 
\end{eqnarray}
for the Fef (YBe) quenching factor for germanium.  As done above, we indicate both, one-sided and two-sided (1 dof) bounds.

Unlike the case of scalar mediators, in the small-mass limit, the bounds for vector mediators are very similar regardless of whether quarks are charged under the new interaction. This is so because in this limit the $Z'$ contribution to the event rates is dominated by the incoherent scattering off electrons, as seen in figure~\ref{fig:events-dresden}. We also notice that for all models, the regions obtained with the YBe quenching factor present a dip around $M_{Z'}\sim 0.03$~MeV and $\left|g_{Z'}\right| \sim 10^{-6}$. This arises because the inclusion of scattering off electrons mediated by such a particle provides a fit which is better than the SM one at $\sim$ 1.5$\sigma$. Comparing the two upper panels, we also see that constraints on the universal model become more stringent than those on the $L_e$ model for $M_{Z'} \gtrsim 0.2$~MeV. For these masses, CE$\nu$NS starts dominating over scattering off electrons.

For sufficiently large mediator masses, the contact-interaction approximation is recovered and the event rates vary with a power of $\left|g_{Z'}\right|/M_{Z'}$, which ranges from a power of two, due the interference with the SM, to a power of four, when the new vector contribution is the dominant one (the interplay between the interference and quadratic contributions is responsible for the disconnected island observed in the upper right excluded region of the $B-L$ model). From the figure, we read that, in this regime, the boundaries of the 90\%~CL excluded regions for the \dresden reactor experiment are approximately described by
\begin{eqnarray}
\frac{\left|g_{Z'}^{\rm univ}\right|}{M_{Z'}/{\rm MeV}} & \gtrsim & 7.5 \; (9.0) \times 10^{-7} ~, \quad {\rm for\;}M_{Z'}\gtrsim 10\;{\rm MeV} ~, \nonumber \\[1.5ex] 
\frac{\left|g_{Z'}^{L_e}\right|}{M_{Z'}/{\rm MeV}} & \gtrsim & 1.7 \; (2.9) \times 10^{-5} ~, \quad {\rm for\;}M_{Z'}\gtrsim 0.1\;{\rm MeV} ~, \\[1.5ex] 
\frac{\left|g_{Z'}^{B-L}\right|}{M_{Z'}/{\rm MeV}} & \gtrsim & \quad \; \; (6.5) \times 10^{-7} ~, \quad {\rm for\;}M_{Z'}\gtrsim 10\;{\rm MeV} ~, \nonumber
\label{eq:vecb2}
\end{eqnarray}
for the Fef (YBe) quenching factor for germanium. For the $B-L$ model, the interplay between scattering off electrons and nucleus and between the interference and quadratic pieces makes the region obtained with the Fef quenching factor not well described by this approximation. Also, as mentioned above, for the universal and $B-L$ models, the combination with COHERENT slightly extends the large-mass part of the excluded region beyond these values as seen in the figure.

\section{Summary and conclusions}
\label{sec:conclusions}

The first evidence of observation of CE$\nu$NS with reactor electron antineutrinos has been recently reported by an experiment conducted at the \dresden reactor site, using the NCC-1701 germanium detector~\cite{Colaresi2022suggestive}. This adds to the previous measurements performed by the COHERENT experiment, which uses neutrinos from a spallation neutron source and has observed CE$\nu$NS using both CsI[Na] and Ar nuclei~\cite{COHERENT:2017ipa, COHERENT:2018imc, COHERENT:2020iec, COHERENT:2020ybo}.

The very low momentum transfer produced by CE$\nu$NS renders this process an excellent probe of new interactions in the neutrino sector. In this paper, we have performed a detailed analysis of the new data from the \dresden experiment, with the aim of deriving powerful constraints on a variety of BSM scenarios. In particular, we have derived new bounds on neutrino NSI with quarks, magnetic moments for electron and muon neutrinos, and several models with light neutral scalar or vector mediators. Our analysis includes the contributions to the event rates from CE$\nu$NS and from elastic scattering off electrons.

In our analysis of the \dresden data, we have taken special care in profiling over the parameters required to accurately model the background, closely following the data release of the experimental collaboration~\cite{Colaresi2022suggestive}. We have also quantified the dependence of the results on the quenching factor by considering two models: one based on the use iron-filtered monochromatic neutrons (Fef) and another one based on photoneutron source measurements (YBe). The impact on the results of the uncertainty on this parameter is approximately indicated by the spread between the bounds obtained for these two models. As for COHERENT, our analysis includes both timing and energy information in the case of CsI, while in the case of the Ar data, we perform a $\chi^2$ analysis using time, energy and F$_{90}$ information. A careful treatment of systematic uncertainties is performed, following the prescriptions provided by the collaboration in refs.~\cite{COHERENT:2018imc, COHERENT:2020ybo}.

We have also quantified the impact of the combination of the new data from the \dresden reactor experiment with COHERENT data. From the phenomenological point of view, the main difference between these two experiments arises from the different flavor composition of the neutrino flux: while a nuclear reactor only emits $\bar\nu_e$, at a spallation neutron source neutrinos are primarily produced from pion decay at rest and therefore, the flux contains an equal admixture of $\nu_\mu$, $\bar\nu_\mu$, and $\nu_e$, with higher energies than the reactor $\bar\nu_e$ flux. Therefore, generically, for models with lepton-flavor dependent effects (such as NSI), COHERENT is a priori sensitive to a larger number of parameters. Nevertheless, the flavor discrimination with COHERENT comes from the timing information and is only partial. Thus, within the current experimental precision, this results in degeneracies among the $\mu$-flavor and $e$-flavor parameters. The new \dresden data bring in complementary information in this respect, as they provide independent constraints on the $e$-flavor parameters. As a result, the combination of both experiments allows breaking (or at least alleviating) degeneracies present in the case of NSI, as illustrated in figure~\ref{fig:nsireg}. Moreover, the combination of data obtained for scattering off different nuclei (Ar and CsI for COHERENT, Ge for the \dresden experiment) also adds additional synergies since, for a given set of NSI parameters, the impact on the weak charge is different depending on the particular nucleus under consideration (see eq.~(\ref{eq:Qalpha-nsi})).

In the case of neutrino magnetic moment and light mediators with masses below ${\cal O} (100\;{\rm MeV})$, we generically find that the \dresden experiment outperforms the bounds obtained for COHERENT (an obvious exception is the magnetic moment of muon neutrinos, which is only constrained by COHERENT, since for the \dresden experiment only a $\bar\nu_e$ flux is available). This is a priori expected, since the \dresden experiment is sensitive to much lower values of the momentum transfer, a critical feature which drives the experimental reach of these scenarios. On the other hand, for mediators with masses above ${\cal O} (100\;{\rm MeV})$, we find that COHERENT and \dresden lead to similar bounds for the universal and $B-L$ scenarios, while if the new interaction is coupled only to leptons, the \dresden bound is more restrictive.

We finish by commenting on the role of ES.  While the event rates from this process are insignificant in the SM, their contribution could be significantly enhanced in the presence of new physics effects. Most notably, such contribution could pass the signal selection cuts for both the \dresden experiment and the COHERENT CsI detector. In fact, for the analysis of COHERENT CsI, this is a novel point of our analysis with respect to previous works in the literature. Our results show that the inclusion of ES is particularly relevant in scenarios with light vector mediators. In the case of COHERENT CsI, this improves the bound on the coupling obtained from CE$\nu$NS alone for the $B-L$ and universal models by at least a factor $\sim 5$, for mediator masses below $\sim 20$~MeV. And a similar improvement is obtained on the bounds derived from the \dresden reactor data when taking ES into account. Conversely, in the reconstructed energy range of the \dresden experiment, the contribution of ES induced by a neutrino magnetic moment and by light scalar mediators is smaller than that from CE$\nu$NS. At higher energies, however, ES would also dominate over CE$\nu$NS at the \dresden reactor experiment. Therefore, extending the reconstructed energy region to higher energies could enhance the reach of the \dresden experiment within these BSM scenarios.

\acknowledgments{ 
We warmly thank Juan Collar for all the help with the new \dresden data release. We also thank Nicola Cargioli for noticing a plotting bug in figure 1. This project has received funding/support from the European Union's Horizon 2020 research and innovation program under the Marie Skłodowska-Curie grant agreement No 860881-HIDDeN, and from grants RTI2018-095979, PID2019-108892RB-I00, PID2019-105614GB-C21, PID2020-113334GB-I00, PID2020-113644GB-I00, CEX2019-000918-M and CEX2020-001007-S, funded by \linebreak MCIN/AEI/10.13039/501100011033 and by ``ERDF A way of making Europe". PC is also supported by Grant RYC2018-024240-I funded by MCIN/AEI/ 10.13039/501100011033 and by “ESF Investing in your Investing in your future''. M.C.G-G is also supported by U.S.A.-NSF grant PHY-1915093, and by AGAUR (Generalitat de Catalunya) grant 2017-SGR-929. LL is supported by the predoctoral training program non-doctoral research personnel of the Department of Education of the Basque Government. LL and FM are also supported by the European Research Council (ERC) under Grant Agreement No. 101039048-GanESS and the Severo Ochoa Program grant CEX2018-000867-S. SPR is also partially supported by the Portuguese FCT (CERN/FIS-PAR/0004/2019).}

\appendix
\section{Binding energies for neutrino scattering off electrons}
\label{app:levels} 

For germanium we use \cite{x-ray}
\begin{equation}
  Z_{\rm eff}^{\rm Ge}(T_e)=\left\{\begin{array}{lcl} 32\;\;
  &&T_e>11.11\;{\rm keV} \\ 30\;\;&&11.11\;{\rm keV}> T_e>1.4146\;{\rm
    keV}\\ 28\;\;&&1.4146\;{\rm keV}> T_e>1.248\;{\rm
    keV}\\ 26\;\;&&1.248\;{\rm keV}> T_e>1.217\;{\rm
    keV}\\ 22\;\;&&1.217\;{\rm keV}> T_e>0.1801\;{\rm
    keV}\\ 20\;\;&&0.1801\;{\rm keV}> T_e>0.1249\;{\rm
    keV}\\ 18\;\;&&0.1249\;{\rm keV}> T_e>0.1208\;{\rm
    keV}\\ 14\;\;&&0.1208\;{\rm keV}> T_e>0.0298\;{\rm
    keV}\\ 4\;\;&&0.0298\;{\rm keV}> T_e \end{array}\right.
\end{equation}  
while for caesium and for iodine we use \cite{x-ray}
\begin{equation}
\begin{array}{cc}
  Z^\mathrm{Cs}_{\rm eff}(T_e)=\left\{\begin{array}{lcl} 55\;\;
  &&T_e>35.99\;{\rm keV} \\ 53\;\;&&35.99\;{\rm keV}> T_e>5.71\;{\rm
    keV}\\ 51\;\;&&5.71\;{\rm keV}> T_e>5.36\;{\rm
    keV}\\ 49\;\;&&5.36\;{\rm keV}> T_e>5.01\;{\rm
    keV}\\ 45\;\;&&5.01\;{\rm keV}> T_e>1.21\;{\rm
    keV}\\ 43\;\;&&1.21\;{\rm keV}> T_e>1.07\;{\rm
    keV}\\ 41\;\;&&1.07\;{\rm keV}> T_e>1\;{\rm keV}\\ 37\;\;&&1\;{\rm
    keV}> T_e>0.74\;{\rm keV}\\ 33\;\;&&0.74\;{\rm keV}>
  T_e>0.73\;{\rm keV}\\ 27\;\;&&0.73\;{\rm keV}> T_e>0.23\;{\rm
    keV}\\ 25\;\;&&0.23\;{\rm keV}> T_e>0.17\;{\rm
    keV}\\ 23\;\;&&0.17\;{\rm keV}> T_e>0.16\;{\rm
    keV}\\ 19\;\;&&0.16\;{\rm keV}> T_e \end{array}\right.  &
  Z^\mathrm{I}_{\rm eff}(T_e)=\left\{\begin{array}{lcl} 53\;\;
  &&T_e>33.17\;{\rm keV} \\ 51\;\;&&33.17\;{\rm keV}> T_e>5.19\;{\rm
    keV}\\ 49\;\;&&5.19\;{\rm keV}> T_e>4.86\;{\rm
    keV}\\ 47\;\;&&4.86\;{\rm keV}> T_e>4.56\;{\rm
    keV}\\ 43\;\;&&4.56\;{\rm keV}> T_e>1.07\;{\rm
    keV}\\ 41\;\;&&1.07\;{\rm keV}> T_e>0.93\;{\rm
    keV}\\ 39\;\;&&0.93\;{\rm keV}> T_e>0.88\;{\rm
    keV}\\ 35\;\;&&0.88\;{\rm keV}> T_e>0.63\;{\rm
    keV}\\ 31\;\;&&0.63\;{\rm keV}> T_e>0.62\;{\rm
    keV}\\ 25\;\;&&0.62\;{\rm keV}> T_e>0.19\;{\rm
    keV}\\ 23\;\;&&0.19\;{\rm keV}> T_e>0.124\;{\rm
    keV}\\ 21\;\;&&0.124\;{\rm keV}> T_e>0.123\;{\rm
    keV}\\ 17\;\;&&0.123\;{\rm keV}> T_e \end{array}\right.
  \end{array}
\end{equation}
noting that $Z_{\rm eff}^\mathrm{CsI}(T_e) = \frac{1}{2} \left[Z_{\rm
    eff}^\mathrm{Cs}(T_e) + Z_{\rm eff}^\mathrm{I}(T_e)\right]$.

\bibliographystyle{JHEP} \bibliography{references}

\providecommand{\href}[2]{#2}\begingroup\raggedright\begin{thebibliography}{100}

\bibitem{Freedman:1973yd}
D.~Z. Freedman, \emph{Coherent effects of a weak neutral current},
  \href{https://doi.org/10.1103/PhysRevD.9.1389}{\emph{Physical Review D}
  {\bfseries 9} (1974) 1389}.

\bibitem{COHERENT:2017ipa}
{\scshape COHERENT} collaboration, D.~Akimov et~al., \emph{{Observation of
  coherent elastic neutrino-nucleus scattering}},
  \href{https://doi.org/10.1126/science.aao0990}{\emph{Science} {\bfseries 357}
  (2017) 1123} [\href{https://arxiv.org/abs/1708.01294}{{\ttfamily
  1708.01294}}].

\bibitem{COHERENT:2018imc}
{\scshape COHERENT} collaboration, D.~Akimov et~al., \emph{{COHERENT
  Collaboration data release from the first observation of coherent elastic
  neutrino-nucleus scattering}},
  \href{https://arxiv.org/abs/1804.09459}{{\ttfamily 1804.09459}}.

\bibitem{COHERENT:2020iec}
{\scshape COHERENT} collaboration, D.~Akimov et~al., \emph{{First measurement
  of coherent elastic neutrino-nucleus scattering on Argon}},
  \href{https://doi.org/10.1103/PhysRevLett.126.012002}{\emph{Phys. Rev. Lett.}
  {\bfseries 126} (2021) 012002}
  [\href{https://arxiv.org/abs/2003.10630}{{\ttfamily 2003.10630}}].

\bibitem{COHERENT:2020ybo}
{\scshape COHERENT} collaboration, D.~Akimov et~al., \emph{{COHERENT
  Collaboration data release from the first detection of coherent elastic
  neutrino-nucleus scattering on Argon}},
  \href{https://arxiv.org/abs/2006.12659}{{\ttfamily 2006.12659}}.

\bibitem{Wong:2004ru}
{\scshape TEXONO} collaboration, H.~T. Wong, \emph{{The TEXONO research program
  on neutrino and astroparticle physics}},
  \href{https://doi.org/10.1142/S0217732304014574}{\emph{Mod. Phys. Lett. A}
  {\bfseries 19} (2004) 1207}.

\bibitem{Belov:2015ufh}
{\scshape \ensuremath{\nu}GeN} collaboration, V.~Belov et~al., \emph{{The
  \ensuremath{\nu}GeN experiment at the Kalinin nuclear power plant}},
  \href{https://doi.org/10.1088/1748-0221/10/12/P12011}{\emph{JINST} {\bfseries
  10} (2015) P12011}.

\bibitem{Aguilar-Arevalo:2016khx}
{\scshape CONNIE} collaboration, A.~Aguilar-Arevalo et~al., \emph{{The CONNIE
  experiment}}, \href{https://doi.org/10.1088/1742-6596/761/1/012057}{\emph{J.
  Phys. Conf. Ser.} {\bfseries 761} (2016) 012057}
  [\href{https://arxiv.org/abs/1608.01565}{{\ttfamily 1608.01565}}].

\bibitem{Agnolet:2016zir}
{\scshape MINER} collaboration, G.~Agnolet et~al., \emph{{Background studies
  for the MINER coherent neutrino scattering reactor experiment}},
  \href{https://doi.org/10.1016/j.nima.2017.02.024}{\emph{Nucl. Instrum. Meth.
  A} {\bfseries 853} (2017) 53}
  [\href{https://arxiv.org/abs/1609.02066}{{\ttfamily 1609.02066}}].

\bibitem{Billard:2016giu}
{\scshape Ricochet} collaboration, J.~Billard et~al., \emph{{Coherent neutrino
  scattering with low temperature bolometers at Chooz reactor complex}},
  \href{https://doi.org/10.1088/1361-6471/aa83d0}{\emph{J. Phys. G} {\bfseries
  44} (2017) 105101} [\href{https://arxiv.org/abs/1612.09035}{{\ttfamily
  1612.09035}}].

\bibitem{Strauss:2017cuu}
{\scshape {$\nu$}-cleus} collaboration, R.~Strauss et~al., \emph{{The
  $\nu$-cleus experiment: a gram-scale fiducial-volume cryogenic detector for
  the first detection of coherent neutrino-nucleus scattering}},
  \href{https://doi.org/10.1140/epjc/s10052-017-5068-2}{\emph{Eur. Phys. J. C}
  {\bfseries 77} (2017) 506}
  [\href{https://arxiv.org/abs/1704.04320}{{\ttfamily 1704.04320}}].

\bibitem{Akimov:2019ogx}
{\scshape RED-100} collaboration, D.~Y. Akimov et~al., \emph{{First
  ground-level laboratory test of the two-phase Xenon emission detector
  RED-100}}, \href{https://doi.org/10.1088/1748-0221/15/02/P02020}{\emph{JINST}
  {\bfseries 15} (2020) P02020}
  [\href{https://arxiv.org/abs/1910.06190}{{\ttfamily 1910.06190}}].

\bibitem{Choi:2020gkm}
J.~J. Choi, \emph{{Neutrino elastic-scattering observation with NaI[Tl]
  (NEON)}}, \href{https://doi.org/10.22323/1.369.0047}{\emph{PoS} {\bfseries
  NuFact2019} (2020) 047}.

\bibitem{CONUS:2020skt}
{\scshape CONUS} collaboration, H.~Bonet et~al., \emph{{Constraints on elastic
  neutrino nucleus scattering in the fully coherent regime from the CONUS
  experiment}},
  \href{https://doi.org/10.1103/PhysRevLett.126.041804}{\emph{Phys. Rev. Lett.}
  {\bfseries 126} (2021) 041804}
  [\href{https://arxiv.org/abs/2011.00210}{{\ttfamily 2011.00210}}].

\bibitem{Colaresi:2021kus}
J.~Colaresi et~al., \emph{{First results from a search for coherent elastic
  neutrino-nucleus scattering at a reactor site}},
  \href{https://doi.org/10.1103/PhysRevD.104.072003}{\emph{Phys. Rev. D}
  {\bfseries 104} (2021) 072003}
  [\href{https://arxiv.org/abs/2108.02880}{{\ttfamily 2108.02880}}].

\bibitem{Colaresi2022suggestive}
J.~Colaresi, J.~I. Collar, T.~W. Hossbach, C.~M. Lewis and K.~M. Yocum,
  \emph{Suggestive evidence for coherent elastic neutrino-nucleus scattering
  from reactor antineutrinos},
  \href{https://arxiv.org/abs/2202.09672}{{\ttfamily 2202.09672}}.

\bibitem{Canas:2018rng}
B.~C. Ca\~nas, E.~A. Garc\'es, O.~G. Miranda and A.~Parada, \emph{{Future
  perspectives for a weak mixing angle measurement in coherent elastic neutrino
  nucleus scattering experiments}},
  \href{https://doi.org/10.1016/j.physletb.2018.07.049}{\emph{Phys. Lett. B}
  {\bfseries 784} (2018) 159}
  [\href{https://arxiv.org/abs/1806.01310}{{\ttfamily 1806.01310}}].

\bibitem{Cadeddu:2018izq}
M.~Cadeddu and F.~Dordei, \emph{{Reinterpreting the weak mixing angle from
  atomic parity violation in view of the Cs neutron rms radius measurement from
  COHERENT}}, \href{https://doi.org/10.1103/PhysRevD.99.033010}{\emph{Phys.
  Rev. D} {\bfseries 99} (2019) 033010}
  [\href{https://arxiv.org/abs/1808.10202}{{\ttfamily 1808.10202}}].

\bibitem{Huang:2019ene}
X.-R. Huang and L.-W. Chen, \emph{{Neutron skin in CsI and low-energy effective
  weak mixing angle from {{COHERENT}} data}},
  \href{https://doi.org/10.1103/PhysRevD.100.071301}{\emph{Phys. Rev. D}
  {\bfseries 100} (2019) 071301}
  [\href{https://arxiv.org/abs/1902.07625}{{\ttfamily 1902.07625}}].

\bibitem{Cadeddu:2019eta}
M.~Cadeddu, F.~Dordei, C.~Giunti, Y.~F. Li and Y.~Y. Zhang, \emph{{Neutrino,
  electroweak, and nuclear physics from COHERENT elastic neutrino-nucleus
  scattering with refined quenching factor}},
  \href{https://doi.org/10.1103/PhysRevD.101.033004}{\emph{Phys. Rev. D}
  {\bfseries 101} (2020) 033004}
  [\href{https://arxiv.org/abs/1908.06045}{{\ttfamily 1908.06045}}].

\bibitem{Cadeddu:2020lky}
M.~Cadeddu, F.~Dordei, C.~Giunti, Y.~F. Li, E.~Picciau and Y.~Y. Zhang,
  \emph{{Physics results from the first COHERENT observation of coherent
  elastic neutrino-nucleus scattering in argon and their combination with
  cesium-iodide data}},
  \href{https://doi.org/10.1103/PhysRevD.102.015030}{\emph{Phys. Rev. D}
  {\bfseries 102} (2020) 015030}
  [\href{https://arxiv.org/abs/2005.01645}{{\ttfamily 2005.01645}}].

\bibitem{Cadeddu:2021ijh}
M.~Cadeddu et~al., \emph{{New insights into nuclear physics and weak mixing
  angle using electroweak probes}},
  \href{https://doi.org/10.1103/PhysRevC.104.065502}{\emph{Phys. Rev. C}
  {\bfseries 104} (2021) 065502}
  [\href{https://arxiv.org/abs/2102.06153}{{\ttfamily 2102.06153}}].

\bibitem{Cadeddu:2017etk}
M.~Cadeddu, C.~Giunti, Y.~F. Li and Y.~Y. Zhang, \emph{{Average CsI neutron
  density distribution from COHERENT data}},
  \href{https://doi.org/10.1103/PhysRevLett.120.072501}{\emph{Phys. Rev. Lett.}
  {\bfseries 120} (2018) 072501}
  [\href{https://arxiv.org/abs/1710.02730}{{\ttfamily 1710.02730}}].

\bibitem{Ciuffoli:2018qem}
E.~Ciuffoli, J.~Evslin, Q.~Fu and J.~Tang, \emph{{Extracting nuclear form
  factors with coherent neutrino scattering}},
  \href{https://doi.org/10.1103/PhysRevD.97.113003}{\emph{Phys. Rev. D}
  {\bfseries 97} (2018) 113003}
  [\href{https://arxiv.org/abs/1801.02166}{{\ttfamily 1801.02166}}].

\bibitem{Papoulias:2019lfi}
D.~K. Papoulias, T.~S. Kosmas, R.~Sahu, V.~K.~B. Kota and M.~Hota,
  \emph{{Constraining nuclear physics parameters with current and future
  {{COHERENT}} data}},
  \href{https://doi.org/10.1016/j.physletb.2019.135133}{\emph{Phys. Lett. B}
  {\bfseries 800} (2020) 135133}
  [\href{https://arxiv.org/abs/1903.03722}{{\ttfamily 1903.03722}}].

\bibitem{Coloma:2020nhf}
P.~Coloma, I.~Esteban, M.~C. Gonzalez-Garcia and J.~Menéndez,
  \emph{{Determining the nuclear neutron distribution from coherent elastic
  neutrino-nucleus scattering: current results and future prospects}},
  \href{https://doi.org/10.1007/JHEP08(2020)030}{\emph{JHEP} {\bfseries 08}
  (2020) 030} [\href{https://arxiv.org/abs/2006.08624}{{\ttfamily
  2006.08624}}].

\bibitem{Barranco:2005yy}
J.~Barranco, O.~G. Miranda and T.~I. Rashba, \emph{{Probing new physics with
  coherent neutrino scattering off nuclei}},
  \href{https://doi.org/10.1088/1126-6708/2005/12/021}{\emph{JHEP} {\bfseries
  12} (2005) 021} [\href{https://arxiv.org/abs/hep-ph/0508299}{{\ttfamily
  hep-ph/0508299}}].

\bibitem{Formaggio:2011jt}
J.~A. Formaggio, E.~Figueroa-Feliciano and A.~J. Anderson, \emph{{Sterile
  neutrinos, coherent scattering and oscillometry measurements with
  low-temperature bolometers}},
  \href{https://doi.org/10.1103/PhysRevD.85.013009}{\emph{Phys. Rev. D}
  {\bfseries 85} (2012) 013009}
  [\href{https://arxiv.org/abs/1107.3512}{{\ttfamily 1107.3512}}].

\bibitem{Anderson:2012pn}
A.~J. Anderson et~al., \emph{{Measuring active-to-sterile neutrino oscillations
  with neutral current coherent neutrino-nucleus scattering}},
  \href{https://doi.org/10.1103/PhysRevD.86.013004}{\emph{Phys. Rev. D}
  {\bfseries 86} (2012) 013004}
  [\href{https://arxiv.org/abs/1201.3805}{{\ttfamily 1201.3805}}].

\bibitem{Dutta:2015nlo}
B.~Dutta, Y.~Gao, R.~Mahapatra, N.~Mirabolfathi, L.~E. Strigari and J.~W.
  Walker, \emph{{Sensitivity to oscillation with a sterile fourth generation
  neutrino from ultra-low threshold neutrino-nucleus coherent scattering}},
  \href{https://doi.org/10.1103/PhysRevD.94.093002}{\emph{Phys. Rev. D}
  {\bfseries 94} (2016) 093002}
  [\href{https://arxiv.org/abs/1511.02834}{{\ttfamily 1511.02834}}].

\bibitem{Cerdeno:2016sfi}
D.~G. Cerde\~no, M.~Fairbairn, T.~Jubb, P.~A.~N. Machado, A.~C. Vincent and
  C.~B\oe{}hm, \emph{{Physics from solar neutrinos in dark matter direct
  detection experiments}},
  \href{https://doi.org/10.1007/JHEP09(2016)048}{\emph{JHEP} {\bfseries 05}
  (2016) 118} [\href{https://arxiv.org/abs/1604.01025}{{\ttfamily
  1604.01025}}].

\bibitem{Dent:2016wcr}
J.~B. Dent, B.~Dutta, S.~Liao, J.~L. Newstead, L.~E. Strigari and J.~W. Walker,
  \emph{{Probing light mediators at ultralow threshold energies with coherent
  elastic neutrino-nucleus scattering}},
  \href{https://doi.org/10.1103/PhysRevD.96.095007}{\emph{Phys. Rev. D}
  {\bfseries 96} (2017) 095007}
  [\href{https://arxiv.org/abs/1612.06350}{{\ttfamily 1612.06350}}].

\bibitem{Coloma:2017egw}
P.~Coloma, P.~B. Denton, M.~C. Gonzalez-Garcia, M.~Maltoni and T.~Schwetz,
  \emph{{Curtailing the dark side in non-standard neutrino interactions}},
  \href{https://doi.org/10.1007/JHEP04(2017)116}{\emph{JHEP} {\bfseries 04}
  (2017) 116} [\href{https://arxiv.org/abs/1701.04828}{{\ttfamily
  1701.04828}}].

\bibitem{Kosmas:2017zbh}
T.~S. Kosmas, D.~K. Papoulias, M.~T\'ortola and J.~W.~F. Valle, \emph{{Probing
  light sterile neutrino signatures at reactor and spallation neutron source
  neutrino experiments}},
  \href{https://doi.org/10.1103/PhysRevD.96.063013}{\emph{Phys. Rev. D}
  {\bfseries 96} (2017) 063013}
  [\href{https://arxiv.org/abs/1703.00054}{{\ttfamily 1703.00054}}].

\bibitem{Ge:2017mcq}
S.-F. Ge and I.~M. Shoemaker, \emph{{Constraining photon portal dark matter
  with {{TEXONO}} and {{COHERENT}} data}},
  \href{https://doi.org/10.1007/JHEP11(2018)066}{\emph{JHEP} {\bfseries 11}
  (2018) 066} [\href{https://arxiv.org/abs/1710.10889}{{\ttfamily
  1710.10889}}].

\bibitem{Shoemaker:2017lzs}
I.~M. Shoemaker, \emph{{COHERENT search strategy for beyond standard model
  neutrino interactions}},
  \href{https://doi.org/10.1103/PhysRevD.95.115028}{\emph{Phys. Rev. D}
  {\bfseries 95} (2017) 115028}
  [\href{https://arxiv.org/abs/1703.05774}{{\ttfamily 1703.05774}}].

\bibitem{Coloma:2017ncl}
P.~Coloma, M.~C. Gonzalez-Garcia, M.~Maltoni and T.~Schwetz, \emph{{COHERENT
  enlightenment of the neutrino dark side}},
  \href{https://doi.org/10.1103/PhysRevD.96.115007}{\emph{Phys. Rev. D}
  {\bfseries 96} (2017) 115007}
  [\href{https://arxiv.org/abs/1708.02899}{{\ttfamily 1708.02899}}].

\bibitem{Liao:2017uzy}
J.~Liao and D.~Marfatia, \emph{{COHERENT constraints on nonstandard neutrino
  interactions}},
  \href{https://doi.org/10.1016/j.physletb.2017.10.046}{\emph{Phys. Lett. B}
  {\bfseries 775} (2017) 54}
  [\href{https://arxiv.org/abs/1708.04255}{{\ttfamily 1708.04255}}].

\bibitem{Canas:2017umu}
B.~C. Ca\~nas, E.~A. Garc\'es, O.~G. Miranda and A.~Parada, \emph{{The reactor
  antineutrino anomaly and low energy threshold neutrino experiments}},
  \href{https://doi.org/10.1016/j.physletb.2017.11.074}{\emph{Phys. Lett. B}
  {\bfseries 776} (2018) 451}
  [\href{https://arxiv.org/abs/1708.09518}{{\ttfamily 1708.09518}}].

\bibitem{Dent:2017mpr}
J.~B. Dent, B.~Dutta, S.~Liao, J.~L. Newstead, L.~E. Strigari and J.~W. Walker,
  \emph{{Accelerator and reactor complementarity in coherent neutrino-nucleus
  scattering}}, \href{https://doi.org/10.1103/PhysRevD.97.035009}{\emph{Phys.
  Rev. D} {\bfseries 97} (2018) 035009}
  [\href{https://arxiv.org/abs/1711.03521}{{\ttfamily 1711.03521}}].

\bibitem{Papoulias:2017qdn}
D.~K. Papoulias and T.~S. Kosmas, \emph{{COHERENT constraints to conventional
  and exotic neutrino physics}},
  \href{https://doi.org/10.1103/PhysRevD.97.033003}{\emph{Phys. Rev. D}
  {\bfseries 97} (2018) 033003}
  [\href{https://arxiv.org/abs/1711.09773}{{\ttfamily 1711.09773}}].

\bibitem{Farzan:2018gtr}
Y.~Farzan, M.~Lindner, W.~Rodejohann and X.-J. Xu, \emph{{Probing neutrino
  coupling to a light scalar with coherent neutrino scattering}},
  \href{https://doi.org/10.1007/JHEP05(2018)066}{\emph{JHEP} {\bfseries 05}
  (2018) 066} [\href{https://arxiv.org/abs/1802.05171}{{\ttfamily
  1802.05171}}].

\bibitem{Billard:2018jnl}
J.~Billard, J.~Johnston and B.~J. Kavanagh, \emph{{Prospects for exploring new
  physics in coherent elastic neutrino-nucleus scattering}},
  \href{https://doi.org/10.1088/1475-7516/2018/11/016}{\emph{JCAP} {\bfseries
  11} (2018) 016} [\href{https://arxiv.org/abs/1805.01798}{{\ttfamily
  1805.01798}}].

\bibitem{Coloma:2019mbs}
P.~Coloma, I.~Esteban, M.~C. Gonzalez-Garcia and M.~Maltoni, \emph{{Improved
  global fit to non-standard neutrino interactions using COHERENT energy and
  timing data}}, \href{https://doi.org/10.1007/JHEP02(2020)023}{\emph{JHEP}
  {\bfseries 02} (2020) 023}
  [\href{https://arxiv.org/abs/1911.09109}{{\ttfamily 1911.09109}}].

\bibitem{Chaves:2021pey}
M.~Chaves and T.~Schwetz, \emph{{Resolving the LMA-dark NSI degeneracy with
  coherent neutrino-nucleus scattering}},
  \href{https://doi.org/10.1007/JHEP05(2021)042}{\emph{JHEP} {\bfseries 05}
  (2021) 042} [\href{https://arxiv.org/abs/2102.11981}{{\ttfamily
  2102.11981}}].

\bibitem{AristizabalSierra:2018eqm}
D.~Aristizabal~Sierra, V.~De~Romeri and N.~Rojas, \emph{{{{COHERENT}} analysis
  of neutrino generalized interactions}},
  \href{https://doi.org/10.1103/PhysRevD.98.075018}{\emph{Phys. Rev. D}
  {\bfseries 98} (2018) 075018}
  [\href{https://arxiv.org/abs/1806.07424}{{\ttfamily 1806.07424}}].

\bibitem{Brdar:2018qqj}
V.~Brdar, W.~Rodejohann and X.-J. Xu, \emph{{Producing a new fermion in
  coherent elastic neutrino-nucleus scattering: from neutrino mass to dark
  matter}}, \href{https://doi.org/10.1007/JHEP12(2018)024}{\emph{JHEP}
  {\bfseries 12} (2018) 024}
  [\href{https://arxiv.org/abs/1810.03626}{{\ttfamily 1810.03626}}].

\bibitem{Cadeddu:2018dux}
M.~Cadeddu, C.~Giunti, K.~A. Kouzakov, Y.~F. Li, A.~I. Studenikin and Y.~Y.
  Zhang, \emph{{Neutrino charge radii from COHERENT elastic neutrino-nucleus
  scattering}}, \href{https://doi.org/10.1103/PhysRevD.98.113010}{\emph{Phys.
  Rev. D} {\bfseries 98} (2018) 113010}
  [\href{https://arxiv.org/abs/1810.05606}{{\ttfamily 1810.05606}}].

\bibitem{Blanco:2019vyp}
C.~Blanco, D.~Hooper and P.~Machado, \emph{{Constraining sterile neutrino
  interpretations of the LSND and MiniBooNE anomalies with coherent neutrino
  scattering experiments}},
  \href{https://doi.org/10.1103/PhysRevD.101.075051}{\emph{Phys. Rev. D}
  {\bfseries 101} (2020) 075051}
  [\href{https://arxiv.org/abs/1901.08094}{{\ttfamily 1901.08094}}].

\bibitem{Dutta:2019eml}
B.~Dutta, S.~Liao, S.~Sinha and L.~E. Strigari, \emph{{Searching for beyond the
  Standard Model physics with COHERENT energy and timing data}},
  \href{https://doi.org/10.1103/PhysRevLett.123.061801}{\emph{Phys. Rev. Lett.}
  {\bfseries 123} (2019) 061801}
  [\href{https://arxiv.org/abs/1903.10666}{{\ttfamily 1903.10666}}].

\bibitem{Miranda:2019wdy}
O.~G. Miranda, D.~K. Papoulias, M.~T\'ortola and J.~W.~F. Valle, \emph{{Probing
  neutrino transition magnetic moments with coherent elastic neutrino-nucleus
  scattering}}, \href{https://doi.org/10.1007/JHEP07(2019)103}{\emph{JHEP}
  {\bfseries 07} (2019) 103}
  [\href{https://arxiv.org/abs/1905.03750}{{\ttfamily 1905.03750}}].

\bibitem{CONNIE:2019swq}
{\scshape CONNIE} collaboration, A.~Aguilar-Arevalo et~al., \emph{{Exploring
  low-energy neutrino physics with the Coherent Neutrino Nucleus Interaction
  Experiment}}, \href{https://doi.org/10.1103/PhysRevD.100.092005}{\emph{Phys.
  Rev. D} {\bfseries 100} (2019) 092005}
  [\href{https://arxiv.org/abs/1906.02200}{{\ttfamily 1906.02200}}].

\bibitem{Dutta:2019nbn}
B.~Dutta, D.~Kim, S.~Liao, J.-C. Park, S.~Shin and L.~E. Strigari, \emph{{Dark
  matter signals from timing spectra at neutrino experiments}},
  \href{https://doi.org/10.1103/PhysRevLett.124.121802}{\emph{Phys. Rev. Lett.}
  {\bfseries 124} (2020) 121802}
  [\href{https://arxiv.org/abs/1906.10745}{{\ttfamily 1906.10745}}].

\bibitem{Papoulias:2019txv}
D.~K. Papoulias, \emph{{COHERENT constraints after the COHERENT-2020 quenching
  factor measurement}},
  \href{https://doi.org/10.1103/PhysRevD.102.113004}{\emph{Phys. Rev. D}
  {\bfseries 102} (2020) 113004}
  [\href{https://arxiv.org/abs/1907.11644}{{\ttfamily 1907.11644}}].

\bibitem{Khan:2019cvi}
A.~N. Khan and W.~Rodejohann, \emph{{New physics from COHERENT data with an
  improved quenching factor}},
  \href{https://doi.org/10.1103/PhysRevD.100.113003}{\emph{Phys. Rev. D}
  {\bfseries 100} (2019) 113003}
  [\href{https://arxiv.org/abs/1907.12444}{{\ttfamily 1907.12444}}].

\bibitem{Giunti:2019xpr}
C.~Giunti, \emph{{General COHERENT constraints on neutrino nonstandard
  interactions}},
  \href{https://doi.org/10.1103/PhysRevD.101.035039}{\emph{Phys. Rev. D}
  {\bfseries 101} (2020) 035039}
  [\href{https://arxiv.org/abs/1909.00466}{{\ttfamily 1909.00466}}].

\bibitem{Baxter:2019mcx}
D.~Baxter et~al., \emph{{Coherent elastic neutrino-nucleus scattering at the
  European Spallation Source}},
  \href{https://doi.org/10.1007/JHEP02(2020)123}{\emph{JHEP} {\bfseries 02}
  (2020) 123} [\href{https://arxiv.org/abs/1911.00762}{{\ttfamily
  1911.00762}}].

\bibitem{Canas:2019fjw}
B.~C. Ca\~nas, E.~A. Garc\'es, O.~G. Miranda, A.~Parada and
  G.~S\'anchez~Garc\'{\i}a, \emph{{Interplay between nonstandard and nuclear
  constraints in coherent elastic neutrino-nucleus scattering experiments}},
  \href{https://doi.org/10.1103/PhysRevD.101.035012}{\emph{Phys. Rev. D}
  {\bfseries 101} (2020) 035012}
  [\href{https://arxiv.org/abs/1911.09831}{{\ttfamily 1911.09831}}].

\bibitem{Miranda:2020zji}
O.~G. Miranda, D.~K. Papoulias, M.~T\'ortola and J.~W.~F. Valle, \emph{{Probing
  new neutral gauge bosons with CE$\nu$NS and neutrino-electron scattering}},
  \href{https://doi.org/10.1103/PhysRevD.101.073005}{\emph{Phys. Rev. D}
  {\bfseries 101} (2020) 073005}
  [\href{https://arxiv.org/abs/2002.01482}{{\ttfamily 2002.01482}}].

\bibitem{Flores:2020lji}
L.~J. Flores, N.~Nath and E.~Peinado, \emph{{Non-standard neutrino interactions
  in U(1)' model after COHERENT data}},
  \href{https://doi.org/10.1007/JHEP06(2020)045}{\emph{JHEP} {\bfseries 06}
  (2020) 045} [\href{https://arxiv.org/abs/2002.12342}{{\ttfamily
  2002.12342}}].

\bibitem{Miranda:2020tif}
O.~G. Miranda, D.~K. Papoulias, G.~S\'anchez~Garc\'{\i}a, O.~Sanders,
  M.~T\'ortola and J.~W.~F. Valle, \emph{{Implications of the first detection
  of coherent elastic neutrino-nucleus scattering (CE$\nu$NS) with Liquid
  Argon}}, \href{https://doi.org/10.1007/JHEP05(2020)130}{\emph{JHEP}
  {\bfseries 05} (2020) 130}
  [\href{https://arxiv.org/abs/2003.12050}{{\ttfamily 2003.12050}}].

\bibitem{Hurtado:2020vlj}
N.~Hurtado, H.~Mir, I.~M. Shoemaker, E.~Welch and J.~Wyenberg, \emph{{Dark
  matter-neutrino interconversion at {{COHERENT}}, direct detection, and the
  early Universe}},
  \href{https://doi.org/10.1103/PhysRevD.102.015006}{\emph{Phys. Rev. D}
  {\bfseries 102} (2020) 015006}
  [\href{https://arxiv.org/abs/2005.13384}{{\ttfamily 2005.13384}}].

\bibitem{Miranda:2020syh}
O.~G. Miranda, D.~K. Papoulias, O.~Sanders, M.~T\'ortola and J.~W.~F. Valle,
  \emph{{Future CE$\nu$NS experiments as probes of lepton unitarity and
  light-sterile neutrinos}},
  \href{https://doi.org/10.1103/PhysRevD.102.113014}{\emph{Phys. Rev. D}
  {\bfseries 102} (2020) 113014}
  [\href{https://arxiv.org/abs/2008.02759}{{\ttfamily 2008.02759}}].

\bibitem{Cadeddu:2020nbr}
M.~Cadeddu et~al., \emph{{Constraints on light vector mediators through
  coherent elastic neutrino nucleus scattering data from COHERENT}},
  \href{https://doi.org/10.1007/JHEP01(2021)116}{\emph{JHEP} {\bfseries 01}
  (2021) 116} [\href{https://arxiv.org/abs/2008.05022}{{\ttfamily
  2008.05022}}].

\bibitem{Shoemaker:2021hvm}
I.~M. Shoemaker and E.~Welch, \emph{{Sailing the CE$\nu$NS seas of non-standard
  neutrino interactions with the coherent CAPTAIN Mills experiment}},
  \href{https://arxiv.org/abs/2103.08401}{{\ttfamily 2103.08401}}.

\bibitem{delaVega:2021wpx}
L.~M.~G. de~la Vega, L.~J. Flores, N.~Nath and E.~Peinado,
  \emph{{Complementarity between dark matter direct searches and
  CE\ensuremath{\nu}NS experiments in U(1)' models}},
  \href{https://doi.org/10.1007/JHEP09(2021)146}{\emph{JHEP} {\bfseries 09}
  (2021) 146} [\href{https://arxiv.org/abs/2107.04037}{{\ttfamily
  2107.04037}}].

\bibitem{Liao:2021yog}
J.~Liao, H.~Liu and D.~Marfatia, \emph{{Coherent neutrino scattering and the
  Migdal effect on the quenching factor}},
  \href{https://doi.org/10.1103/PhysRevD.104.015005}{\emph{Phys. Rev. D}
  {\bfseries 104} (2021) 015005}
  [\href{https://arxiv.org/abs/2104.01811}{{\ttfamily 2104.01811}}].

\bibitem{CONUS:2021dwh}
{\scshape CONUS} collaboration, H.~Bonet et~al., \emph{{Novel constraints on
  neutrino physics beyond the standard model from the CONUS experiment}},
  \href{https://doi.org/10.1007/JHEP05(2022)085}{\emph{JHEP} {\bfseries 05}
  (2022) 085} [\href{https://arxiv.org/abs/2110.02174}{{\ttfamily
  2110.02174}}].

\bibitem{Flores:2021kzl}
L.~J. Flores, N.~Nath and E.~Peinado, \emph{{CE\ensuremath{\nu}NS as a probe of
  flavored generalized neutrino interactions}},
  \href{https://doi.org/10.1103/PhysRevD.105.055010}{\emph{Phys. Rev. D}
  {\bfseries 105} (2022) 055010}
  [\href{https://arxiv.org/abs/2112.05103}{{\ttfamily 2112.05103}}].

\bibitem{Li:2022jfl}
Y.-F. Li and S.-y. Xia, \emph{{Constraining light mediators via detection of
  coherent elastic solar neutrino nucleus scattering}},
  \href{https://doi.org/10.1016/j.nuclphysb.2022.115737}{\emph{Nucl. Phys. B}
  {\bfseries 977} (2022) 115737}
  [\href{https://arxiv.org/abs/2201.05015}{{\ttfamily 2201.05015}}].

\bibitem{AristizabalSierra:2019zmy}
D.~Aristizabal~Sierra, J.~Liao and D.~Marfatia, \emph{{Impact of form factor
  uncertainties on interpretations of coherent elastic neutrino-nucleus
  scattering data}}, \href{https://doi.org/10.1007/JHEP06(2019)141}{\emph{JHEP}
  {\bfseries 06} (2019) 141}
  [\href{https://arxiv.org/abs/1902.07398}{{\ttfamily 1902.07398}}].

\bibitem{Abdullah:2020iiv}
M.~Abdullah, D.~Aristizabal~Sierra, B.~Dutta and L.~E. Strigari,
  \emph{{Coherent elastic neutrino-nucleus scattering with directional
  detectors}}, \href{https://doi.org/10.1103/PhysRevD.102.015009}{\emph{Phys.
  Rev. D} {\bfseries 102} (2020) 015009}
  [\href{https://arxiv.org/abs/2003.11510}{{\ttfamily 2003.11510}}].

\bibitem{Fernandez-Moroni:2021nap}
G.~Fernandez-Moroni et~al., \emph{{The physics potential of a reactor neutrino
  experiment with Skipper-CCDs: searching for new physics with light
  mediators}}, \href{https://doi.org/10.1007/JHEP02(2022)127}{\emph{JHEP}
  {\bfseries 02} (2022) 127}
  [\href{https://arxiv.org/abs/2108.07310}{{\ttfamily 2108.07310}}].

\bibitem{Bertuzzo:2021opb}
E.~Bertuzzo, G.~Grilli~di Cortona and L.~M.~D. Ramos, \emph{{Probing light
  vector mediators with coherent scattering at future facilities}},
  \href{https://doi.org/10.1007/JHEP06(2022)075}{\emph{JHEP} {\bfseries 06}
  (2022) 075} [\href{https://arxiv.org/abs/2112.04020}{{\ttfamily
  2112.04020}}].

\bibitem{Bonet:2022imz}
H.~Bonet et~al., \emph{{First limits on neutrino electromagnetic properties
  from the CONUS experiment}},
  \href{https://arxiv.org/abs/2201.12257}{{\ttfamily 2201.12257}}.

\bibitem{Lindner:2016wff}
M.~Lindner, W.~Rodejohann and X.-J. Xu, \emph{{Coherent neutrino-nucleus
  scattering and new neutrino interactions}},
  \href{https://doi.org/10.1007/JHEP03(2017)097}{\emph{JHEP} {\bfseries 03}
  (2017) 097} [\href{https://arxiv.org/abs/1612.04150}{{\ttfamily
  1612.04150}}].

\bibitem{Erler:2004in}
J.~Erler and M.~J. Ramsey-Musolf, \emph{{The weak mixing angle at low
  energies}}, \href{https://doi.org/10.1103/PhysRevD.72.073003}{\emph{Phys.
  Rev. D} {\bfseries 72} (2005) 073003}
  [\href{https://arxiv.org/abs/hep-ph/0409169}{{\ttfamily hep-ph/0409169}}].

\bibitem{Erler:2017knj}
J.~Erler and R.~Ferro-Hern\'andez, \emph{{Weak mixing angle in the Thomson
  limit}}, \href{https://doi.org/10.1007/JHEP03(2018)196}{\emph{JHEP}
  {\bfseries 03} (2018) 196}
  [\href{https://arxiv.org/abs/1712.09146}{{\ttfamily 1712.09146}}].

\bibitem{Klein:1999qj}
S.~Klein and J.~Nystrand, \emph{{Exclusive vector meson production in
  relativistic heavy ion collisions}},
  \href{https://doi.org/10.1103/PhysRevC.60.014903}{\emph{Phys. Rev. C}
  {\bfseries 60} (1999) 014903}
  [\href{https://arxiv.org/abs/hep-ph/9902259}{{\ttfamily hep-ph/9902259}}].

\bibitem{Helm:1956zz}
R.~H. Helm, \emph{{Inelastic and elastic scattering of 187-{MeV} electrons from
  selected even-even nuclei}},
  \href{https://doi.org/10.1103/PhysRev.104.1466}{\emph{Phys. Rev.} {\bfseries
  104} (1956) 1466}.

\bibitem{Klos:2013rwa}
P.~Klos, J.~Men\'endez, D.~Gazit and A.~Schwenk, \emph{{Large-scale nuclear
  structure calculations for spin-dependent WIMP scattering with chiral
  effective field theory currents}},
  \href{https://doi.org/10.1103/PhysRevD.88.083516}{\emph{Phys. Rev. D}
  {\bfseries 88} (2013) 083516}
  [\href{https://arxiv.org/abs/1304.7684}{{\ttfamily 1304.7684}}].

\bibitem{Hoferichter:2018acd}
M.~Hoferichter, P.~Klos, J.~Men\'endez and A.~Schwenk, \emph{{Nuclear structure
  factors for general spin-independent WIMP-nucleus scattering}},
  \href{https://doi.org/10.1103/PhysRevD.99.055031}{\emph{Phys. Rev. D}
  {\bfseries 99} (2019) 055031}
  [\href{https://arxiv.org/abs/1812.05617}{{\ttfamily 1812.05617}}].

\bibitem{Hoferichter:2016nvd}
M.~Hoferichter, P.~Klos, J.~Men\'endez and A.~Schwenk, \emph{{Analysis
  strategies for general spin-independent WIMP-nucleus scattering}},
  \href{https://doi.org/10.1103/PhysRevD.94.063505}{\emph{Phys. Rev. D}
  {\bfseries 94} (2016) 063505}
  [\href{https://arxiv.org/abs/1605.08043}{{\ttfamily 1605.08043}}].

\bibitem{ParticleDataGroup:2020ssz}
{\scshape Particle Data Group} collaboration, P.~A. Zyla et~al., \emph{{Review
  of Particle Physics}},
  \href{https://doi.org/10.1093/ptep/ptaa104}{\emph{PTEP} {\bfseries 2020}
  (2020) 083C01}.

\bibitem{Angeli:2013epw}
I.~Angeli and K.~P. Marinova, \emph{{Table of experimental nuclear ground state
  charge radii: An update}},
  \href{https://doi.org/10.1016/j.adt.2011.12.006}{\emph{Atom. Data Nucl. Data
  Tabl.} {\bfseries 99} (2013) 69}.

\bibitem{Payne:2019wvy}
C.~G. Payne, S.~Bacca, G.~Hagen, W.~Jiang and T.~Papenbrock, \emph{{Coherent
  elastic neutrino-nucleus scattering on $^{40}$Ar from first principles}},
  \href{https://doi.org/10.1103/PhysRevC.100.061304}{\emph{Phys. Rev. C}
  {\bfseries 100} (2019) 061304}
  [\href{https://arxiv.org/abs/1908.09739}{{\ttfamily 1908.09739}}].

\bibitem{Vogel:1989iv}
P.~Vogel and J.~Engel, \emph{{Neutrino electromagnetic form-factors}},
  \href{https://doi.org/10.1103/PhysRevD.39.3378}{\emph{Phys. Rev. D}
  {\bfseries 39} (1989) 3378}.

\bibitem{Tomalak:2019ibg}
O.~Tomalak and R.~J. Hill, \emph{{Theory of elastic neutrino-electron
  scattering}}, \href{https://doi.org/10.1103/PhysRevD.101.033006}{\emph{Phys.
  Rev. D} {\bfseries 101} (2020) 033006}
  [\href{https://arxiv.org/abs/1907.03379}{{\ttfamily 1907.03379}}].

\bibitem{Mikaelyan:2002nv}
L.~A. Mikaelyan, \emph{Investigation of neutrino properties in experiments at
  nuclear reactors: present status and prospects},
  \href{https://doi.org/10.1134/1.1495017}{\emph{Phys.Atom.Nucl.} {\bfseries
  65} (2002) 1173} [\href{https://arxiv.org/abs/hep-ph/0210047}{{\ttfamily
  hep-ph/0210047}}].

\bibitem{Scholberg:2005qs}
K.~Scholberg, \emph{{Prospects for measuring coherent neutrino-nucleus elastic
  scattering at a stopped-pion neutrino source}},
  \href{https://doi.org/10.1103/PhysRevD.73.033005}{\emph{Phys. Rev. D}
  {\bfseries 73} (2006) 033005}
  [\href{https://arxiv.org/abs/hep-ex/0511042}{{\ttfamily hep-ex/0511042}}].

\bibitem{Hoferichter:2020osn}
M.~Hoferichter, J.~Men\'endez and A.~Schwenk, \emph{{Coherent elastic
  neutrino-nucleus scattering: EFT analysis and nuclear responses}},
  \href{https://doi.org/10.1103/PhysRevD.102.074018}{\emph{Phys. Rev. D}
  {\bfseries 102} (2020) 074018}
  [\href{https://arxiv.org/abs/2007.08529}{{\ttfamily 2007.08529}}].

\bibitem{Rodejohann:2017vup}
W.~Rodejohann, X.-J. Xu and C.~E. Yaguna, \emph{{Distinguishing between Dirac
  and Majorana neutrinos in the presence of general interactions}},
  \href{https://doi.org/10.1007/JHEP05(2017)024}{\emph{JHEP} {\bfseries 05}
  (2017) 024} [\href{https://arxiv.org/abs/1702.05721}{{\ttfamily
  1702.05721}}].

\bibitem{Shifman:1978zn}
M.~A. Shifman, A.~I. Vainshtein and V.~I. Zakharov, \emph{{Remarks on Higgs
  boson interactions with nucleons}},
  \href{https://doi.org/10.1016/0370-2693(78)90481-1}{\emph{Phys. Lett. B}
  {\bfseries 78} (1978) 443}.

\bibitem{DelNobile:2013sia}
M.~Cirelli, E.~Del~Nobile and P.~Panci, \emph{Tools for model-independent
  bounds in direct dark matter searches},
  \href{https://doi.org/10.1088/1475-7516/2013/10/019}{\emph{JCAP} {\bfseries
  10} (2013) 019} [\href{https://arxiv.org/abs/1307.5955}{{\ttfamily
  1307.5955}}].

\bibitem{Ellis:2018dmb}
J.~Ellis, N.~Nagata and K.~A. Olive, \emph{{Uncertainties in WIMP dark matter
  scattering revisited}},
  \href{https://doi.org/10.1140/epjc/s10052-018-6047-y}{\emph{Eur. Phys. J. C}
  {\bfseries 78} (2018) 569}
  [\href{https://arxiv.org/abs/1805.09795}{{\ttfamily 1805.09795}}].

\bibitem{Huber:2011wv}
P.~Huber, \emph{{On the determination of anti-neutrino spectra from nuclear
  reactors}}, \href{https://doi.org/10.1103/PhysRevC.85.029901}{\emph{Phys.
  Rev. C} {\bfseries 84} (2011) 024617}
  [\href{https://arxiv.org/abs/1106.0687}{{\ttfamily 1106.0687}}].

\bibitem{Mueller:2011nm}
T.~A. Mueller et~al., \emph{{Improved predictions of reactor antineutrino
  spectra}}, \href{https://doi.org/10.1103/PhysRevC.83.054615}{\emph{Phys. Rev.
  C} {\bfseries 83} (2011) 054615}
  [\href{https://arxiv.org/abs/1101.2663}{{\ttfamily 1101.2663}}].

\bibitem{Qian:2018wid}
X.~Qian and J.-C. Peng, \emph{{Physics with reactor neutrinos}},
  \href{https://doi.org/10.1088/1361-6633/aae881}{\emph{Rept. Prog. Phys.}
  {\bfseries 82} (2019) 036201}
  [\href{https://arxiv.org/abs/1801.05386}{{\ttfamily 1801.05386}}].

\bibitem{SuperCDMS:2015eex}
{\scshape SuperCDMS} collaboration, R.~Agnese et~al., \emph{{New results from
  the search for low-mass weakly interacting massive particles with the CDMS
  low ionization threshold experiment}},
  \href{https://doi.org/10.1103/PhysRevLett.116.071301}{\emph{Phys. Rev. Lett.}
  {\bfseries 116} (2016) 071301}
  [\href{https://arxiv.org/abs/1509.02448}{{\ttfamily 1509.02448}}].

\bibitem{Firestone1996}
R.~B. Firestone and V.~S. Shirley. New York: Wiley, 1996.

\bibitem{Antman1966272}
S.~O.~W. Antman, D.~A. Landis and R.~H. Pehl, \emph{Measurements of the {F}ano
  factor and the energy per hole-electron pair in {G}ermanium},
  \href{https://doi.org/https://doi.org/10.1016/0029-554X(66)90386-7}{\emph{Nucl.
  Instrum. Methods} {\bfseries 40} (1966) 272}.

\bibitem{Wei:2016xbw}
W.~Z. Wei, L.~Wang and D.~M. Mei, \emph{{Average energy expended per e-h pair
  for {{Germanium-based}} dark matter experiments}},
  \href{https://doi.org/10.1088/1748-0221/12/04/P04022}{\emph{JINST} {\bfseries
  12} (2017) P04022} [\href{https://arxiv.org/abs/1602.08005}{{\ttfamily
  1602.08005}}].

\bibitem{Collar:2021fcl}
J.~I. Collar, A.~R.~L. Kavner and C.~M. Lewis, \emph{{Germanium response to
  sub-keV nuclear recoils: a multipronged experimental characterization}},
  \href{https://doi.org/10.1103/PhysRevD.103.122003}{\emph{Phys. Rev. D}
  {\bfseries 103} (2021) 122003}
  [\href{https://arxiv.org/abs/2102.10089}{{\ttfamily 2102.10089}}].

\bibitem{Collar:2019ihs}
J.~I. Collar, A.~R.~L. Kavner and C.~M. Lewis, \emph{{Response of CsI[Na] to
  nuclear recoils: impact on coherent elastic neutrino-nucleus scattering
  (CE$\nu$NS)}}, \href{https://doi.org/10.1103/PhysRevD.100.033003}{\emph{Phys.
  Rev. D} {\bfseries 100} (2019) 033003}
  [\href{https://arxiv.org/abs/1907.04828}{{\ttfamily 1907.04828}}].

\bibitem{Creus:2013sau}
W.~Creus, \emph{{Light yield in liquid Argon for dark matter detection}}, Ph.D.
  thesis, Zurich U., 2013.
\newblock 10.5167/uzh-86639.

\bibitem{Esteban:2018ppq}
I.~Esteban, M.~C. Gonzalez-Garcia, M.~Maltoni, I.~Martinez-Soler and
  J.~Salvado, \emph{{Updated constraints on non-standard interactions from
  global analysis of oscillation data}},
  \href{https://doi.org/10.1007/JHEP08(2018)180}{\emph{JHEP} {\bfseries 08}
  (2018) 180} [\href{https://arxiv.org/abs/1805.04530}{{\ttfamily
  1805.04530}}].

\bibitem{Beda:2012zz}
{\scshape GEMMA} collaboration, A.~G. Beda et~al., \emph{{The results of search
  for the neutrino magnetic moment in GEMMA experiment}},
  \href{https://doi.org/10.1155/2012/350150}{\emph{Adv. High Energy Phys.}
  {\bfseries 2012} (2012) 350150}.

\bibitem{Beda:2013mta}
{\scshape GEMMA} collaboration, A.~G. Beda et~al., \emph{{Gemma experiment: the
  results of neutrino magnetic moment search}},
  \href{https://doi.org/10.1134/S1547477113020027}{\emph{Phys. Part. Nucl.
  Lett.} {\bfseries 10} (2013) 139}.

\bibitem{x-ray}
A.~Thompson et~al., \emph{{X-ray data booklet}},
  {\emph{\url{https://xdb.lbl.gov/}} (2009) }.

\end{thebibliography}\endgroup

\end{document}